\documentclass[myepj-spec]{mySvjour}
\usepackage{graphicx}
\usepackage{hyperref}
\usepackage{color}
\usepackage{xspace}
\usepackage{natbib}   
\usepackage{verbatim}
\usepackage{amsmath,amssymb,amsfonts} 
\usepackage{tabularx}

\usepackage{lmodern,bm}                
\usepackage[T1]{sansmath} 
\SetMathAlphabet{\mathsfbf}{sans}{\sansmathencoding}{\sfdefault}{bx}{sl}
\usepackage{etoolbox}
\AtBeginEnvironment{sansmath}{}{}{}

\newcommand{\figsize}{\columnwidth}
\newcommand{\fighspace}{0.4cm}

\definecolor{darkblue1}{rgb}{0,0,.2}
\definecolor{darkblue}{rgb}{0,0,.2}
\definecolor{darkred}{rgb}{0.5,0,0}
\pagecolor{white} 
\color{black}     
\hypersetup{breaklinks=true, 
            colorlinks=true, 
            linkcolor=darkblue1, 
            menucolor=darkblue1, 
            urlcolor=darkblue1,
            citecolor=darkblue1,
            pdftitle={},
            pdfauthor={},
            pdfsubject={},
            pdfkeywords={},
            pdfproducer={}
}
\bibstyle{plain}
\mathchardef\Upsilon="7107
\def\Y#1S{\ensuremath{\Upsilon{(#1S)}}\xspace}

\newcommand{\mZ}{\ensuremath{m_Z}\xspace}
\newcommand{\as}{\ensuremath{\alpha_{\scriptscriptstyle S}}\xspace}

\newcommand{\asZ}{\ensuremath{\as(\mZ^2)}\xspace}

\newcommand{\aZ}{\ensuremath{\alpha(m_Z^2)}\xspace}
\newcommand{\dahadZ}{\ensuremath{\Delta\alpha_{\rm had}(m_Z^2)}\xspace}

\newcommand{\Kbar    }{\kern 0.2em\overline{\kern -0.2em K}{}\xspace}
\newcommand{\Dbar    }{\kern 0.2em\overline{\kern -0.2em D}{}\xspace}
\newcommand{\pbar    }{\kern 0.1em\overline{\kern -0.1em p}{}\xspace}
\newcommand{\nbar    }{\kern 0.1em\overline{\kern -0.1em n}{}\xspace}

\newcommand{\Kz      }{\ensuremath{K^0}\xspace}
\newcommand{\Kzb     }{\ensuremath{\Kbar^0}\xspace}
\newcommand{\KzKzb   }{\ensuremath{\Kz \kern -0.16em \Kzb}\xspace}
\newcommand{\Kp      }{\ensuremath{K^+}\xspace}
\newcommand{\Km      }{\ensuremath{K^-}\xspace}

\newcommand{\KpKm    }{\ensuremath{\Kp \kern -0.16em \Km}\xspace}
\newcommand{\KS      }{\ensuremath{K^0_{\scriptscriptstyle S}}\xspace} 
\newcommand{\KL      }{\ensuremath{K^0_{\scriptscriptstyle L}}\xspace}

\newcommand{\nut}{\ensuremath{\nu_\tau}\xspace}

\newcommand{\pip}{\ensuremath{\pi^+}\xspace}
\newcommand{\pim}{\ensuremath{\pi^-}\xspace}
\newcommand{\piz}{\ensuremath{\pi^0}\xspace}

\newcommand{\ee}{\ensuremath{e^+e^-}\xspace}

\newcommand{\pp}{\ensuremath{\pi^+\pi^-}\xspace}
\newcommand{\pipi}{\ensuremath{\pi^+\pi^-}\xspace}
\newcommand{\pipiz}{\ensuremath{\pi^-\pi^0}\xspace}

\newcommand{\ppg}{\ensuremath{\pi^+\pi^-(\gamma)}\xspace}

\newcommand{\tev}{\ensuremath{\mathrm{\,Te\kern -0.1em V}}\xspace}
\newcommand{\gev}{\ensuremath{\mathrm{\,Ge\kern -0.1em V}}\xspace}
\newcommand{\mev}{\ensuremath{\mathrm{\,Me\kern -0.1em V}}\xspace}
\newcommand{\kev}{\ensuremath{\mathrm{\,ke\kern -0.1em V}}\xspace}
\newcommand{\ev}{\ensuremath{\mathrm{\,e\kern -0.1em V}}\xspace}
\newcommand{\gevc}{\ensuremath{{\mathrm{\,Ge\kern -0.1em V\!/}c}}\xspace}
\newcommand{\mevc}{\ensuremath{{\mathrm{\,Me\kern -0.1em V\!/}c}}\xspace}
\newcommand{\gevcc}{\ensuremath{{\mathrm{\,Ge\kern -0.1em V\!/}c^2}}\xspace}
\newcommand{\mevcc}{\ensuremath{{\mathrm{\,Me\kern -0.1em V\!/}c^2}}\xspace}

\newcommand{\bei}{\begin{itemize}}
\newcommand{\eei}{\end{itemize}}
\newcommand{\ben}{\begin{enumerate}}
\newcommand{\een}{\end{enumerate}}
\newcommand{\beq}{\begin{equation}}
\newcommand{\eeq}{\end{equation}}
\newcommand{\beqn}{\begin{eqnarray}}
\newcommand{\eeqn}{\end{eqnarray}}
\newcommand{\beqns}{\begin{eqnarray*}}
\newcommand{\eeqns}{\end{eqnarray*}}

\newcommand{\amu}{\ensuremath{a_\mu}\xspace}

\newcommand{\amuhadLO}{\ensuremath{\amu^{\rm had,LO}}\xspace}

\newcommand{\amuhadLBL}{\ensuremath{\amu^{\rm had,LBL}}\xspace}
\newcommand{\amuSM}{\ensuremath{\amu^{\rm SM}}\xspace}
\newcommand{\amuExp}{\ensuremath{\amu^{\rm exp}}\xspace}
\newcommand\cf{{cf.}\xspace}
\newcommand{\ea}{{\em et al.}\xspace}
\def\@citex[#1]#2{\if@filesw\immediate\write\@auxout{\string\citation{#2}}\fi
  \@tempcnta\z@\@tempcntb\m@ne\def\@citea{}\@cite{\@for\@citeb:=#2\do
    {\@ifundefined
       {b@\@citeb}{\@citeo\@tempcntb\m@ne\@citea
        \def\@citea{,\penalty\@m\ }{\bf ?}\@warning
       {Citation `\@citeb' on page \thepage \space undefined}}%
    {\setbox\z@\hbox{\global\@tempcntc0\csname b@\@citeb\endcsname\relax}%
     \ifnum\@tempcntc=\z@ \@citeo\@tempcntb\m@ne
       \@citea\def\@citea{,\penalty\@m}
       \hbox{\csname b@\@citeb\endcsname}%
     \else
      \advance\@tempcntb\@ne
      \ifnum\@tempcntb=\@tempcntc
      \else\advance\@tempcntb\m@ne\@citeo
      \@tempcnta\@tempcntc\@tempcntb\@tempcntc\fi\fi}}\@citeo}{#1}}

\def\@citeo{\ifnum\@tempcnta>\@tempcntb\else\@citea
  \def\@citea{,\penalty\@m}%
  \ifnum\@tempcnta=\@tempcntb\the\@tempcnta\else
   {\advance\@tempcnta\@ne\ifnum\@tempcnta=\@tempcntb \else
\def\@citea{--}\fi
    \advance\@tempcnta\m@ne\the\@tempcnta\@citea\the\@tempcntb}\fi\fi}
\newenvironment{myquote}
               {\list{}{\leftmargin0cm\indent}%
                \item\relax}
               {\endlist}
\newcommand\allFontSize{\footnotesize}
\newcommand\detailsSize{\allFontSize}
{\begin{myquote}\detailsSize}{\end{myquote}}

\begin{document}
 
\twocolumn[{%
  \begin{@twocolumnfalse}

    \begin{flushright}
      \normalsize
      \today
    \end{flushright}

    \vspace{-2cm}

    \title{\Large\boldmath Reevaluation of the hadronic vacuum polarisation contributions to the Standard Model predictions of the muon $g-2$ and \sansmath$\mathbf{\boldsymbol\alpha(m_Z^2)}$ using newest hadronic cross-section data}

    \author{M.~Davier\inst{1} \and 
        A.~Hoecker\inst{2} \and 
        B.~Malaescu\inst{3} \and 
        Z.~Zhang\inst{1}}
 
    \institute{Laboratoire de l'Acc{\'e}l{\'e}rateur Lin{\'e}aire,
          IN2P3-CNRS et Universit\'e Paris-Sud 11, F--91898, Orsay Cedex, France \and
          CERN, CH--1211, Geneva 23, Switzerland \and
          Laboratoire de Physique Nucl\'eaire et des Hautes Energies, 
          IN2P3-CNRS et Universit\'es Pierre-et-Marie-Curie et Denis-Diderot, 
          F--75252 Paris Cedex 05, France }
          
    \abstract{
      We reevaluate the hadronic vacuum polarisation contributions to the 
      muon magnetic anomaly and to the running of the electromagnetic coupling 
      constant at the $Z$-boson mass. We include newest $\ee \to {\rm hadrons}$ 
      cross-section data (among others) from the BABAR and VEPP-2000 experiments. 
      For the muon $(g-2)/2$ we find for the lowest-order hadronic contribution  
      $(693.1 \pm 3.4)\cdot10^{-10}$, improving the precision of our previous 
      evaluation by 21\%. The full Standard Model prediction  differs by 
      $3.5\,\sigma$ from the experimental value. The five-quark hadronic 
      contribution to \aZ is evaluated to be $(276.0\pm0.9)\cdot10^{-4}$. 
    }

    \maketitle
  \end{@twocolumnfalse}
}]

\section{~Introduction}
\label{sec:Introduction}

The Standard Model (SM) predictions of the anomalous magnetic moment of the muon, 
$\amu=(g_\mu-2)/2$, with $g_\mu$ the muon gyromagnetic factor, 
and of the running electromagnetic coupling constant, $\alpha(s)$,  a crucial ingredient 
of electroweak theory, are 
limited in precision by    hadronic vacuum polarisation (HVP) contributions. 
The dominant hadronic terms can be calculated with a combination of experimental 
cross-section data, involving \ee annihilation to hadrons, and perturbative QCD. They are 
used to evaluate  energy-squared dispersion integrals ranging from the $\piz\gamma$ 
threshold to infinity. The  kernels occurring in these integrals 
emphasise low photon virtualities, owing to the $1/s$ descent of the cross section, 
and, in case of \amu, to an additional $1/s$ suppression. In the latter case, about 73\% 
of the lowest order hadronic contribution and 59\% of the total uncertainty-squared
are given by the $\ppg$ final state,\footnote
{Throughout this paper, final state photon radiation is implied for all 
   hadronic final states.
} 
while this channel amounts to only 12\% of the hadronic contribution to $\alpha(s)$ 
at $s=\mZ^2$.

In this work, we reevaluate the lowest-order hadronic contribution, \amuhadLO, to 
the muon magnetic anomaly, and the hadronic contribution, \dahadZ, to the running 
\aZ at the $Z$-boson mass using newest $\ee \to {\rm hadrons}$ cross-section 
data. The BABAR Collaboration has essentially completed a programme of precise 
measurements of exclusive hadronic cross sections for all the dominant channels from 
threshold to an energy of 3--5$\;$GeV using the initial-state radiation (ISR) method. 
Also new results are being produced at the VEPP-2000 facility in Novosibirsk, Russia 
in the 1--2$\;$GeV energy range. 
The new data complement the available information on exclusive channels allowing 
to alleviate the need for estimating missing channels with the use of isospin symmetry.  

We reevaluate all the experimental contributions using the software package
HVPTools~\cite{g209}, and add to these narrow resonance contributions evaluated 
analytically and continuum contributions obtained from perturbative QCD.

\section{~Input data}

Exclusive bare hadronic cross-section  measurements are integrated up to 1.8$\;$GeV over the 
relevant dispersion kernels. In the present work 39  channels 
are included, as compared to only 22 in our latest work from 2011~\cite{dhmz2011}.
Thanks to the new measurements only very few final states remain to be estimated using 
isospin symmetry. In the energy range 1.8--3.7$\;$GeV and above 5$\;$GeV 
four-loop perturbative QCD is used~\cite{baikov}. The contributions from the open charm
pair production region between 3.7 and 5$\;$GeV are again computed using experimental data. 
For the narrow resonances $J/\psi$ and $\psi(2S)$ Breit-Wigner line shapes are integrated 
using their currently best known parameters. 

The integration of data points belonging to different experiments with their own
data densities requires a careful treatment especially with respect to  
correlated systematic uncertainties within the same experiment and between
different experiments. Quadratic interpolation of adjacent data
points is performed for each experiment and a local combination between the
interpolations is computed in bins of 1$\;$MeV. Full covariance matrices are 
constructed between experiments and channels. Uncertainties are propagated using
pseudo-experiment generation and  closure tests with known distributions are performed to 
validate both the combination and integration. Where results from different experiments are 
locally inconsistent the combined uncertainty is rescaled according to the local
$\chi^2$ value following the well-known PDG approach. At present, for the dominant 
$\pi^+\pi^-$ channel such inconsistencies are limiting the 
precision of the combination. In most exclusive channels the largest weight 
in the combination is provided by  BABAR measurements.

The following channel-wise discussion focuses on the HVP contribution to $\amu$ as 
it stronger relies on the low-energy experimental data. We mainly explore the impact of the  
data released since our last publication~\cite{dhmz2011}, which provides  references 
to all the older datasets used in the combination.

\begin{figure*}[p]
\begin{center}
\includegraphics[width=130mm]{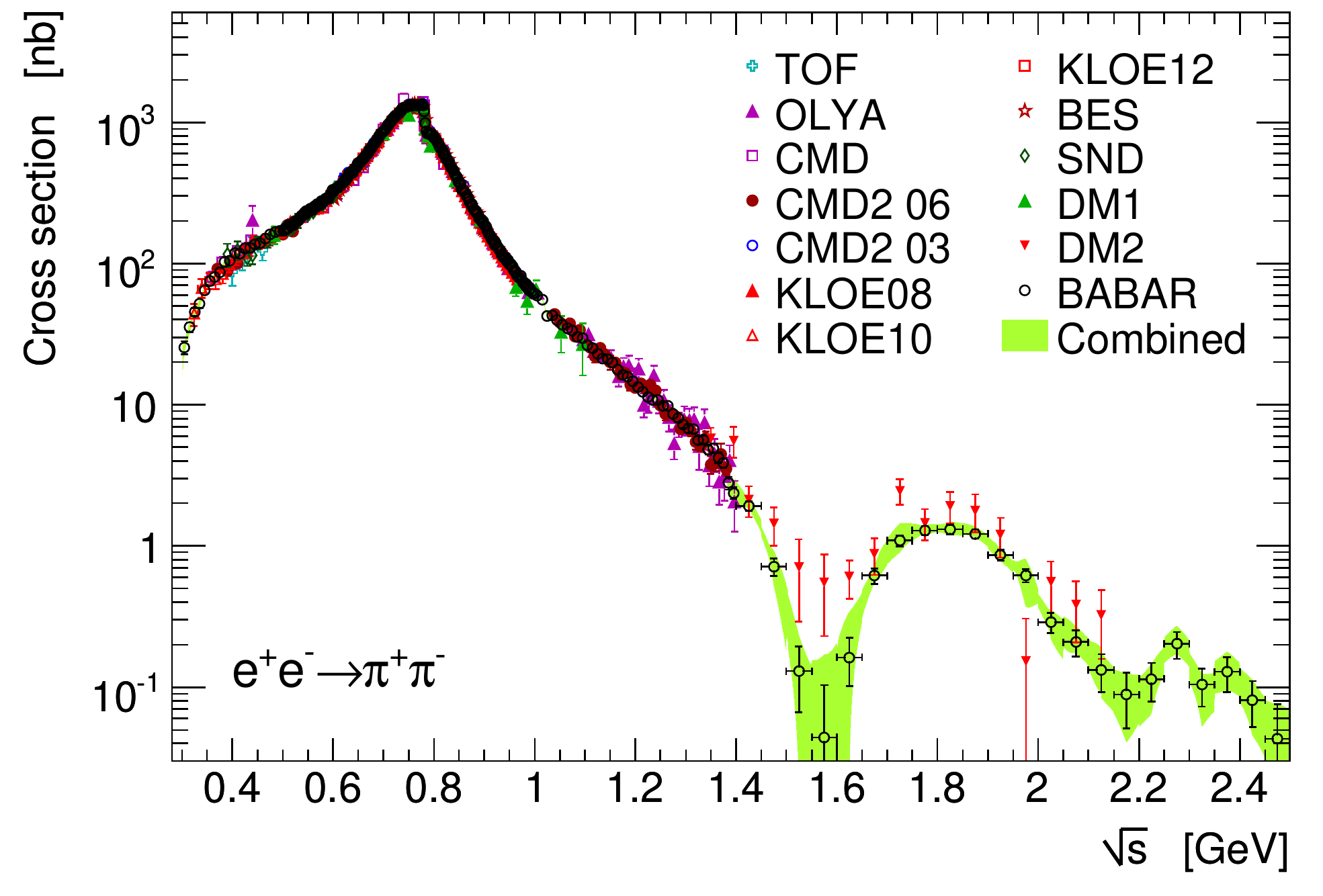}
\vspace{0.5cm}

\includegraphics[width=\figsize]{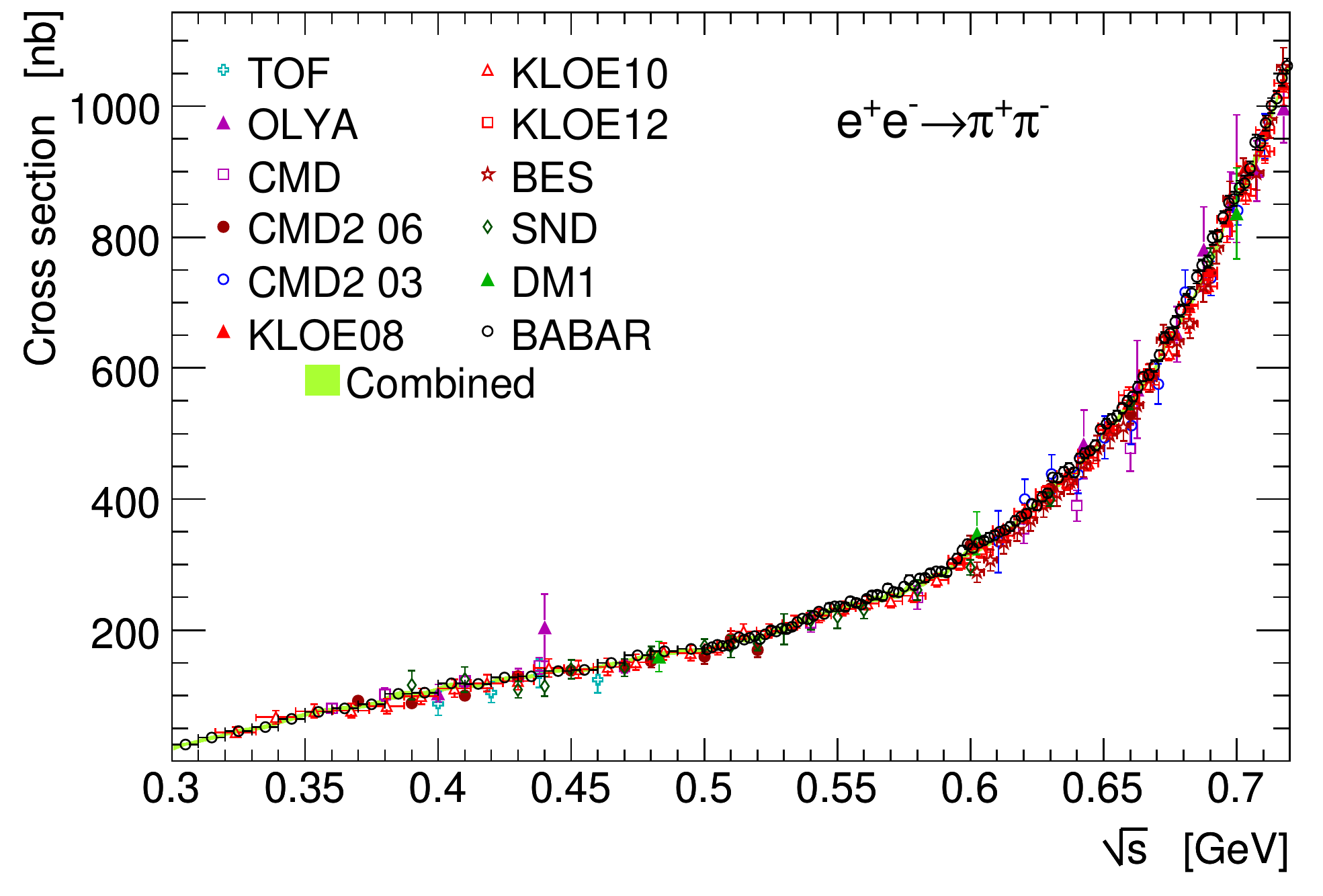}\hspace{\fighspace}
\includegraphics[width=\figsize]{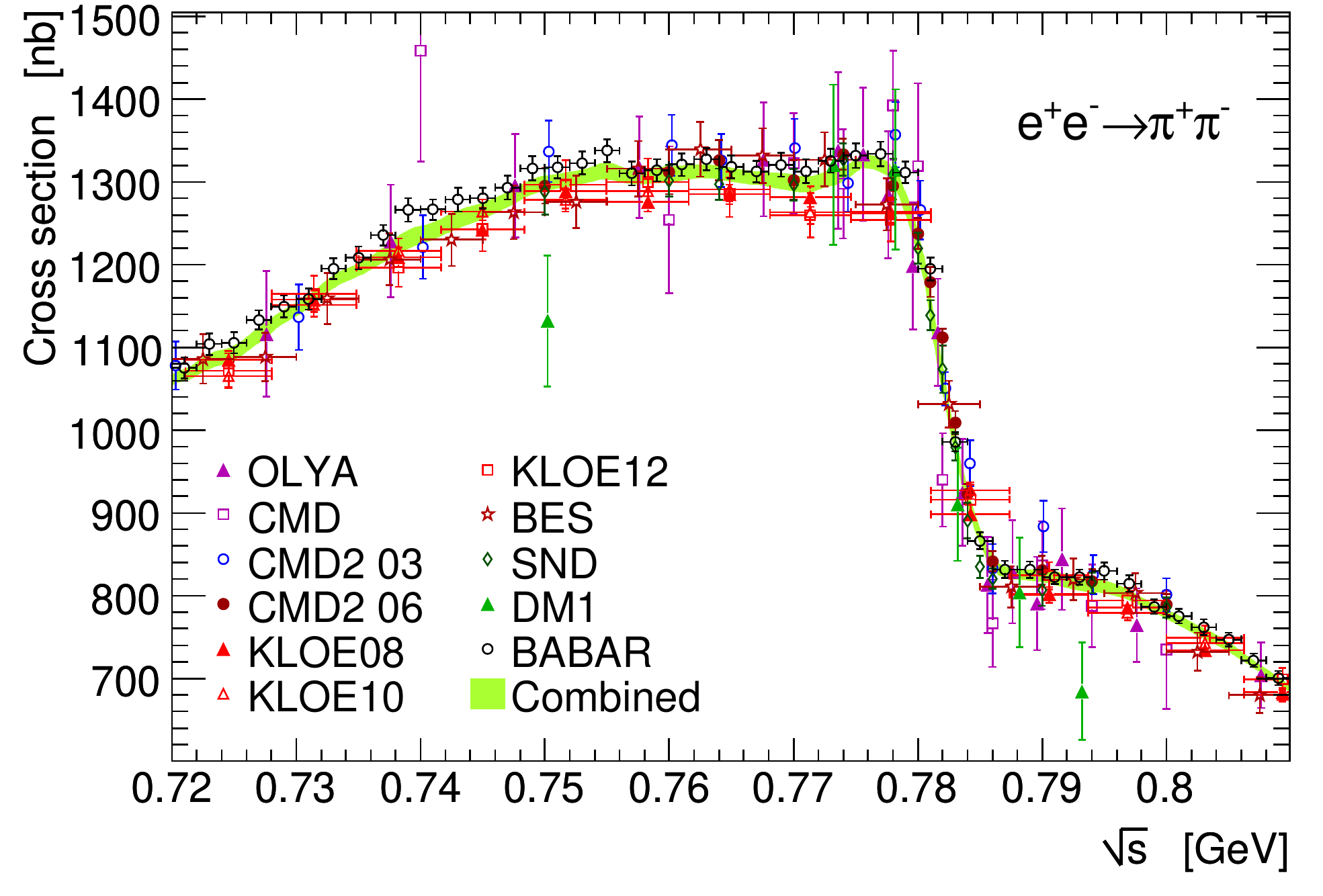}
\vspace{0.2cm}

\includegraphics[width=\figsize]{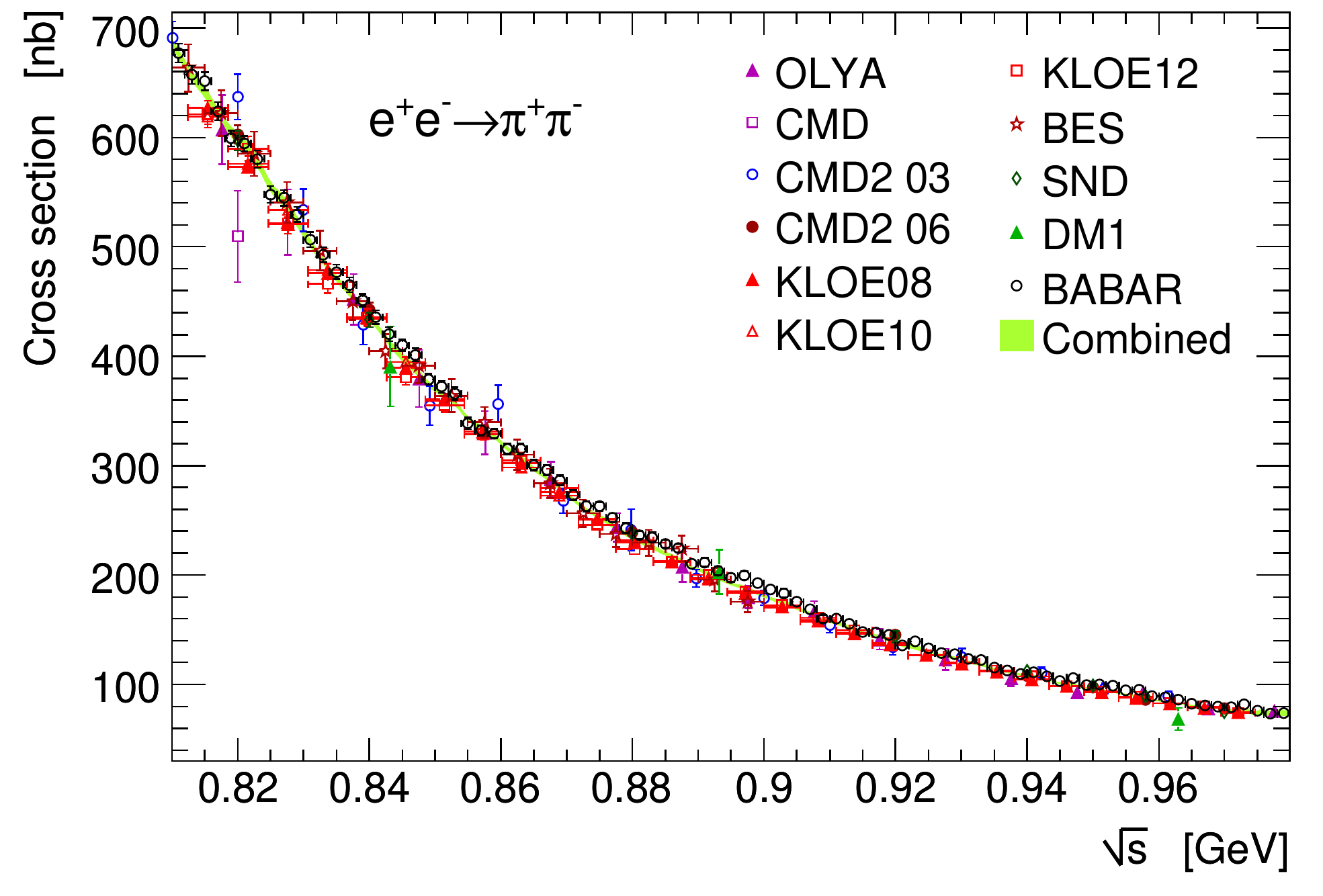}\hspace{\fighspace}
\includegraphics[width=\figsize]{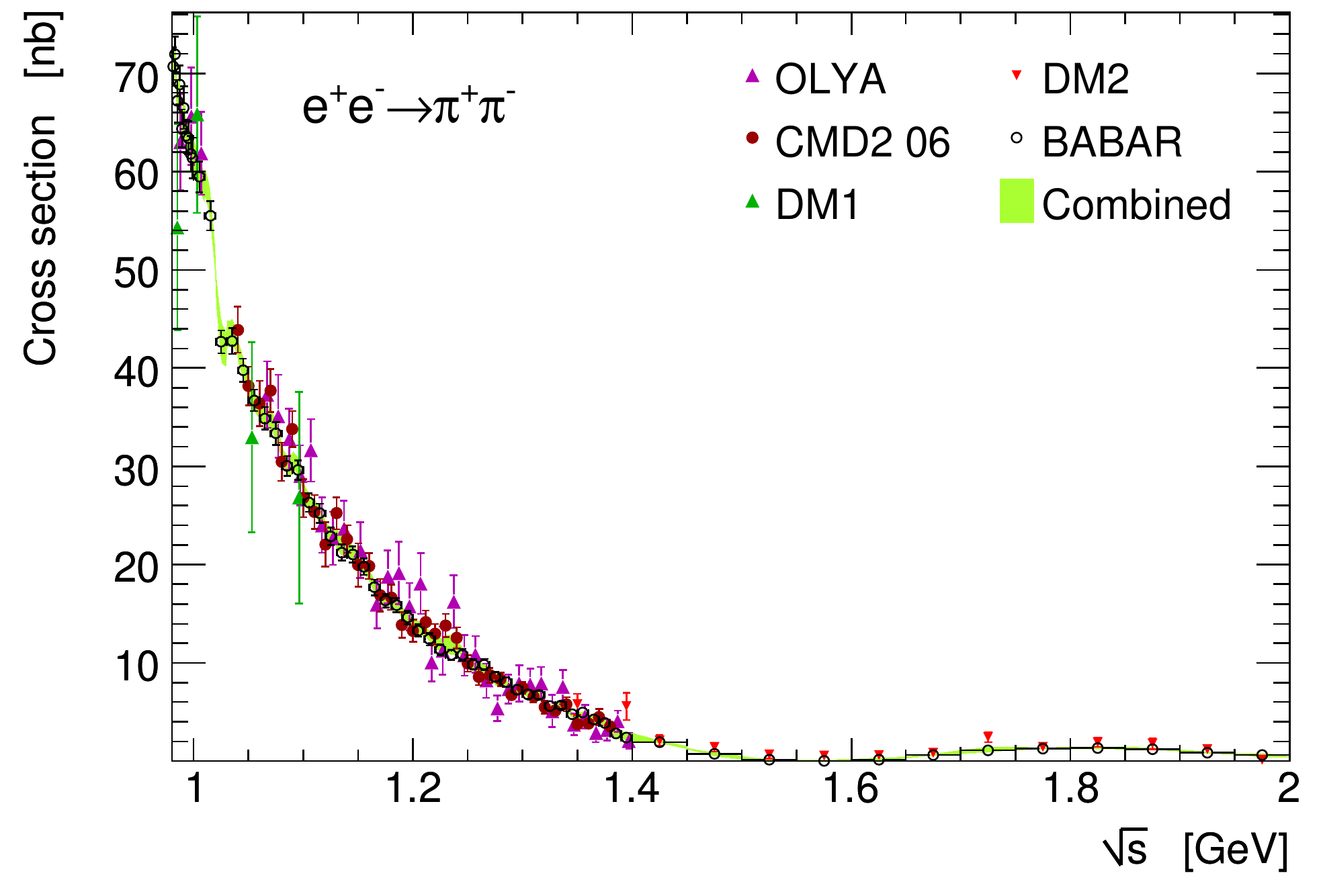}
\end{center}
\vspace{-0.2cm}
\caption[.]{ 
            Bare cross section of $\ee\to\pp$ versus centre-of-mass energy for different 
            energy ranges.  The error bars of the data points include statistical and systematic 
            uncertainties added in quadrature. The green band shows
            the HVPTools combination within its $1\,\sigma$ uncertainty. 
    
}
\label{fig:pipiall}
\end{figure*}
\begin{figure*}[t]
\begin{center}
\includegraphics[width=\figsize]{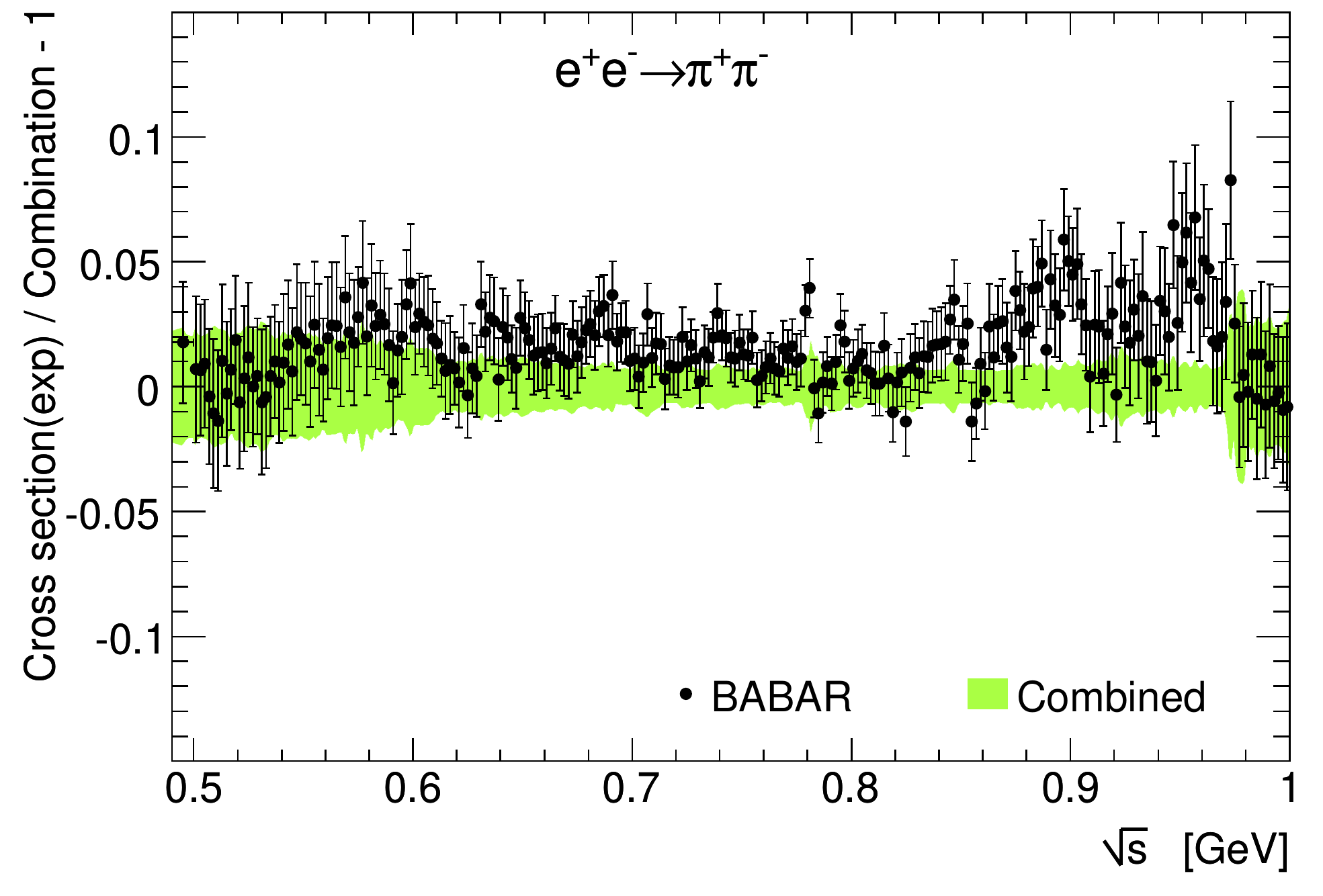}\hspace{\fighspace}
\includegraphics[width=\figsize]{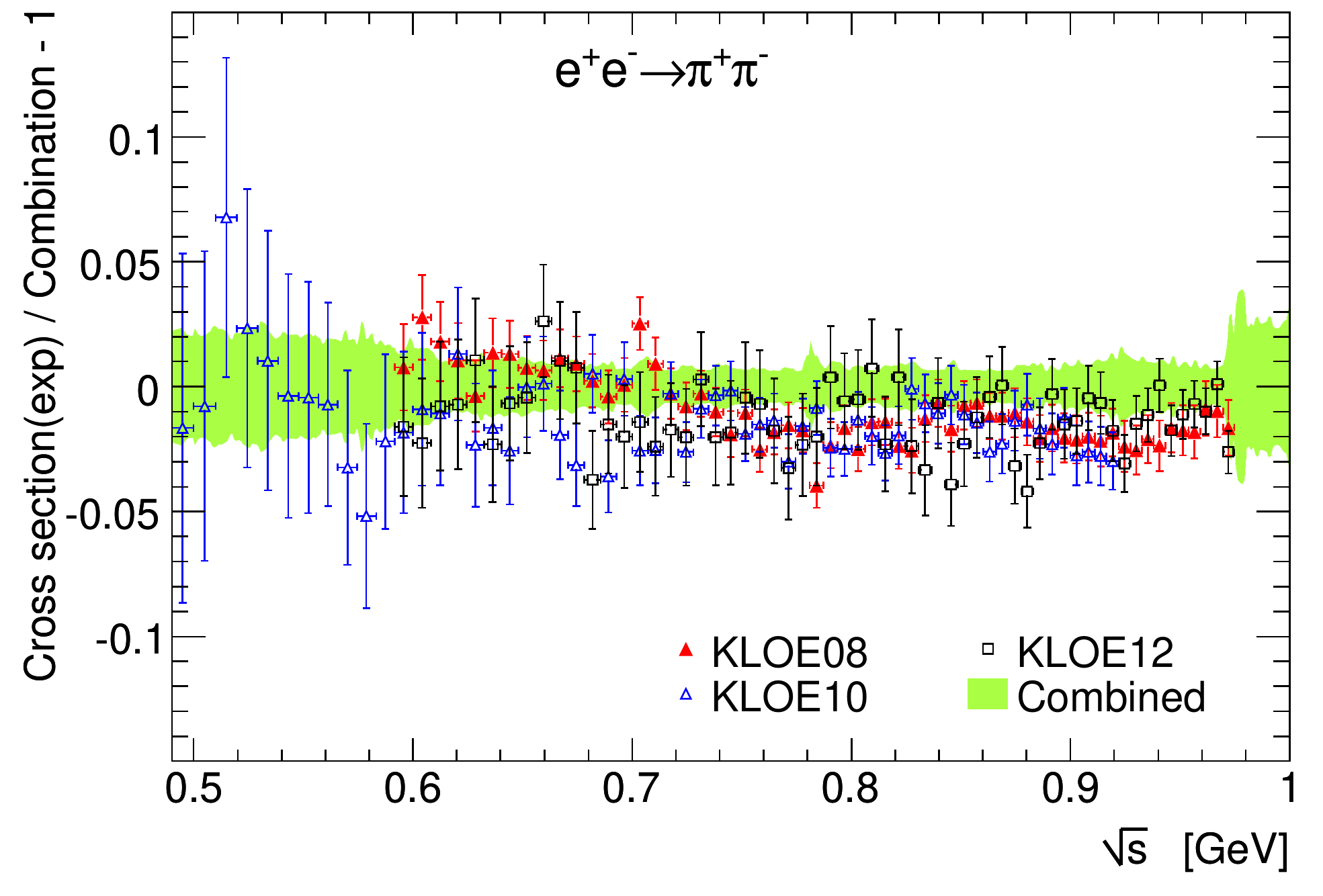}
\vspace{0.2cm}

\includegraphics[width=\figsize]{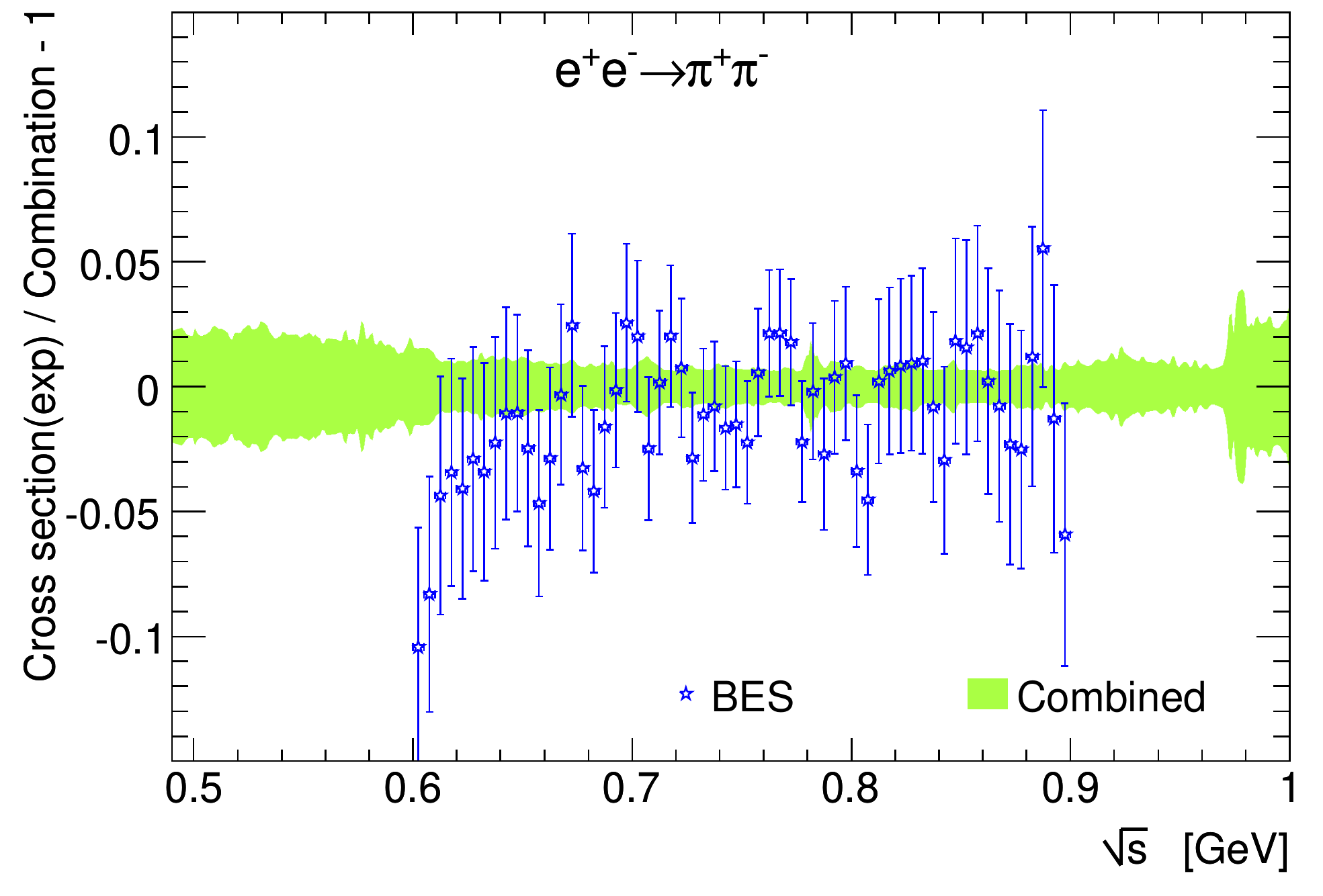}\hspace{\fighspace}
\includegraphics[width=\figsize]{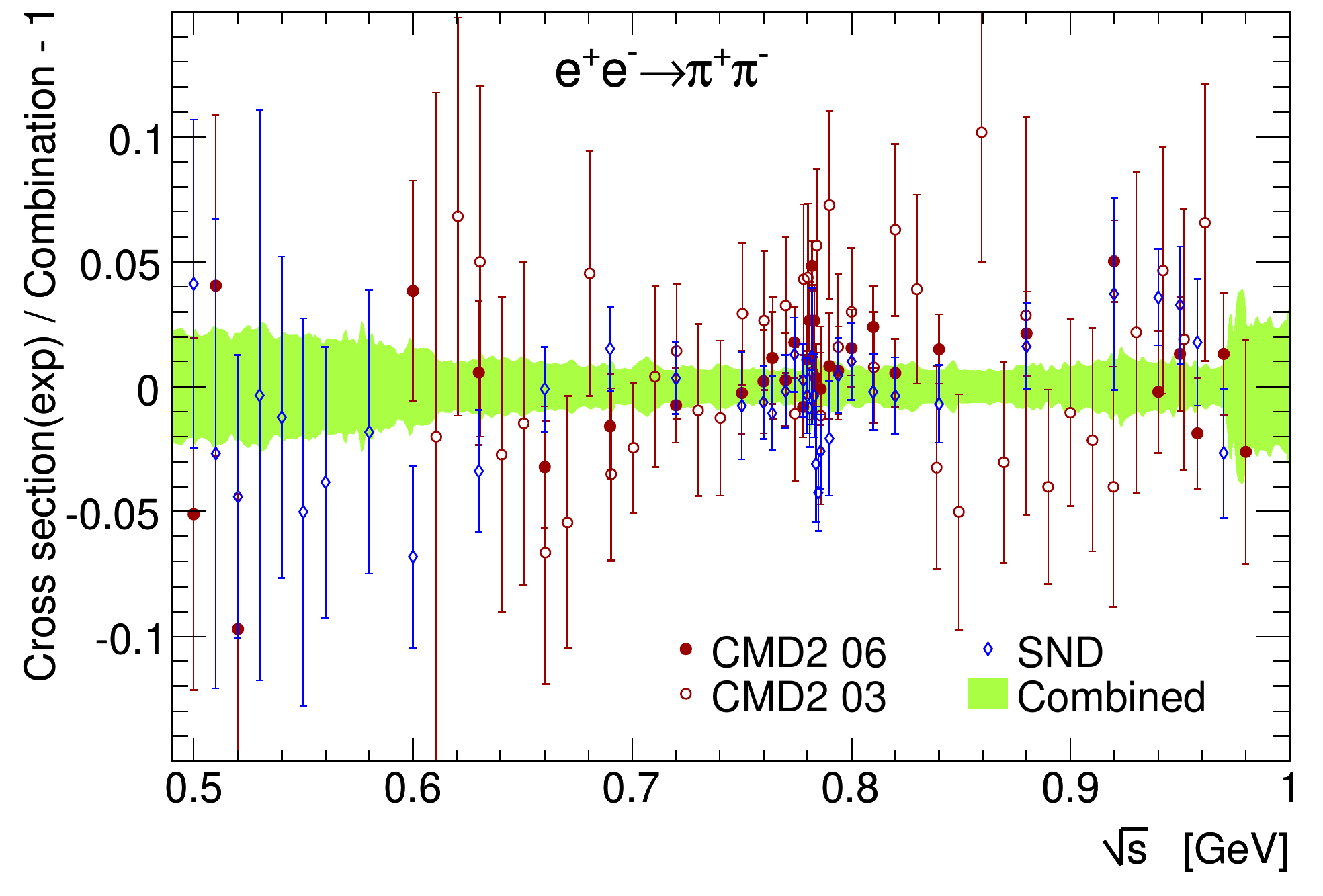}
\end{center}
\vspace{-0.2cm}
\caption[.]{ 
            Comparison between individual $\ee\to\pp$ cross-section measurements from 
            BABAR~\cite{babarpipi1,babarpipi2}, KLOE\,08~\cite{kloe08}, KLOE\,10~\cite{kloe10},
            KLOE\,12~\cite{kloe12}, BESIII~\cite{bes2015},
            CMD2\,03~\cite{cmd203}, CMD2\,06~\cite{cmd2new}, SND~\cite{snd2pi}, and 
            the HVPTools combination. The error bars include statistical and systematic 
            uncertainties added in quadrature. 
}
\label{fig:comppipi}
\end{figure*}
\begin{figure*}[t]
\includegraphics[width=\figsize]{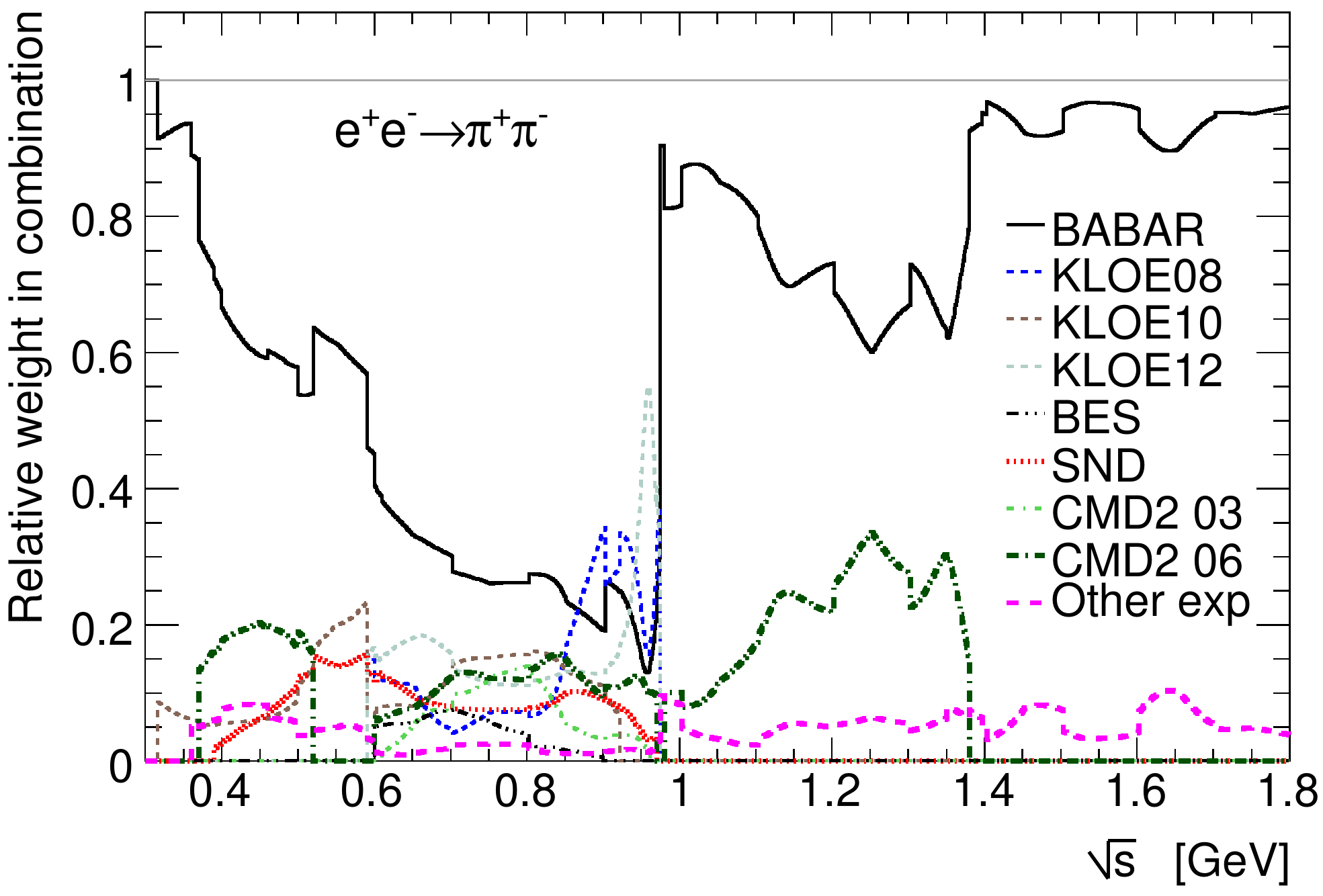}\hspace{\fighspace}
\includegraphics[width=\figsize]{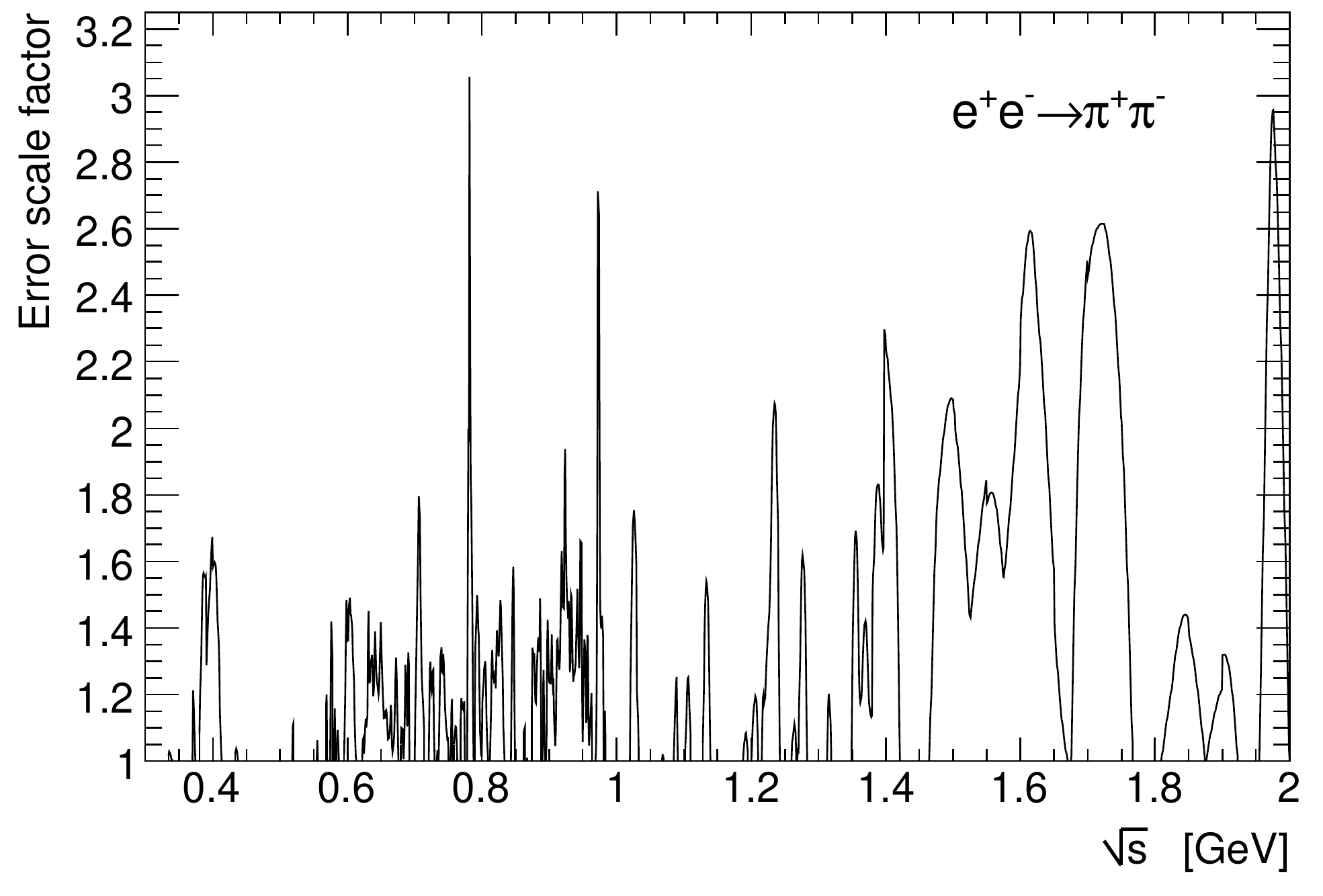}
\vspace{0.1cm}

\caption[.]{ 
            Left: relative local weight per experiment contributing to the $\ee\to\pp$
            cross-section combination versus centre-of-mass energy. 
            Right: local scale factor versus centre-of-mass energy applied 
            to the combined \pp cross-section uncertainty to account for inconsistency 
            in the individual measurements. }
\label{fig:weights}
\end{figure*}
\begin{figure}[t]
\begin{center}
\includegraphics[width=\figsize]{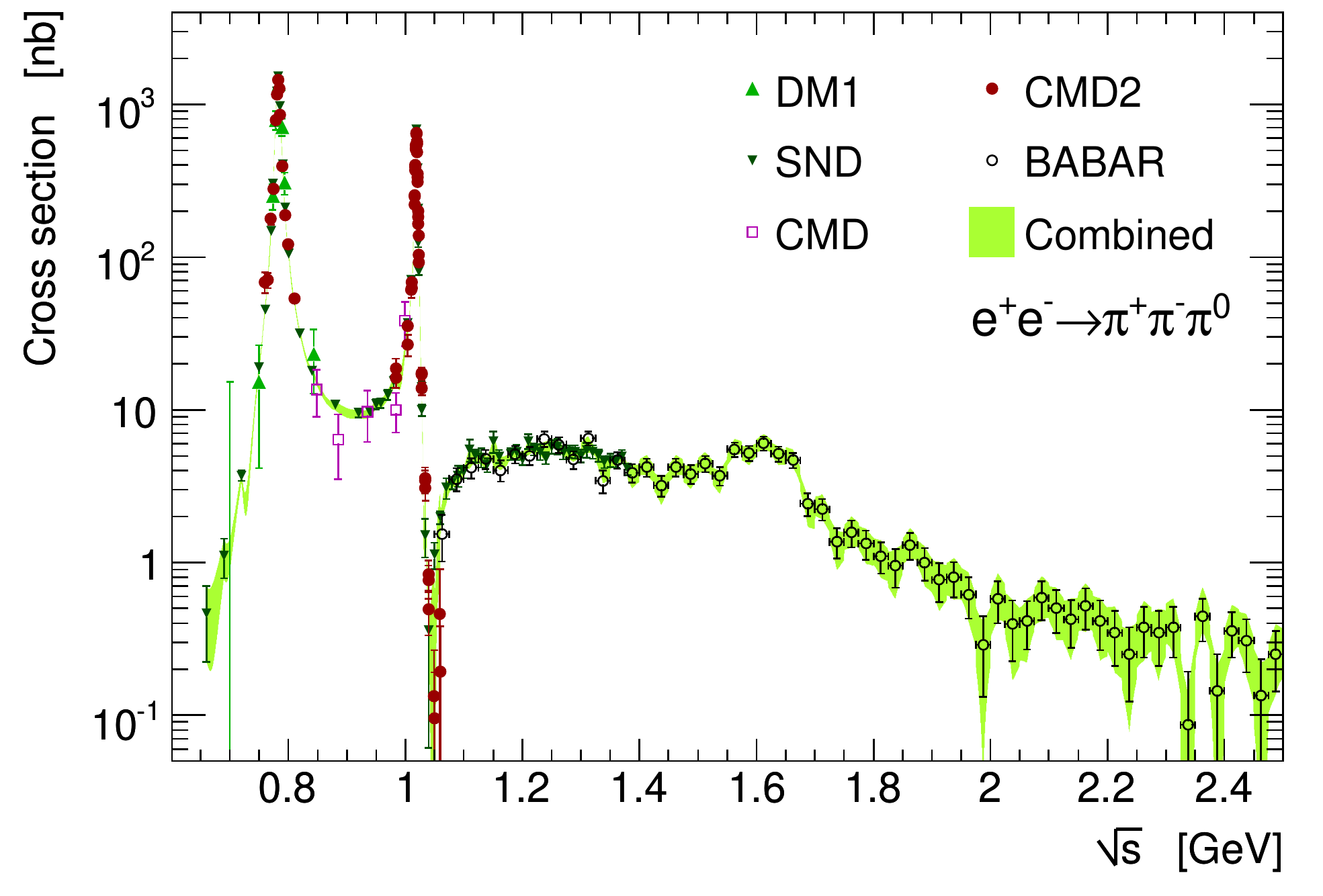}
\end{center}
\vspace{-0.2cm}
\caption[.]{ 
            Bare cross section of $\ee\to\pp\piz$ versus centre-of-mass energy.
            The error bars of the data points include statistical and systematic 
            uncertainties added in quadrature. The green band shows
            the HVPTools combination within its $1\,\sigma$ uncertainty. 
}
\label{fig:xsec3pi}
\end{figure}
\begin{figure*}[t]
\begin{center}
\includegraphics[width=\figsize]{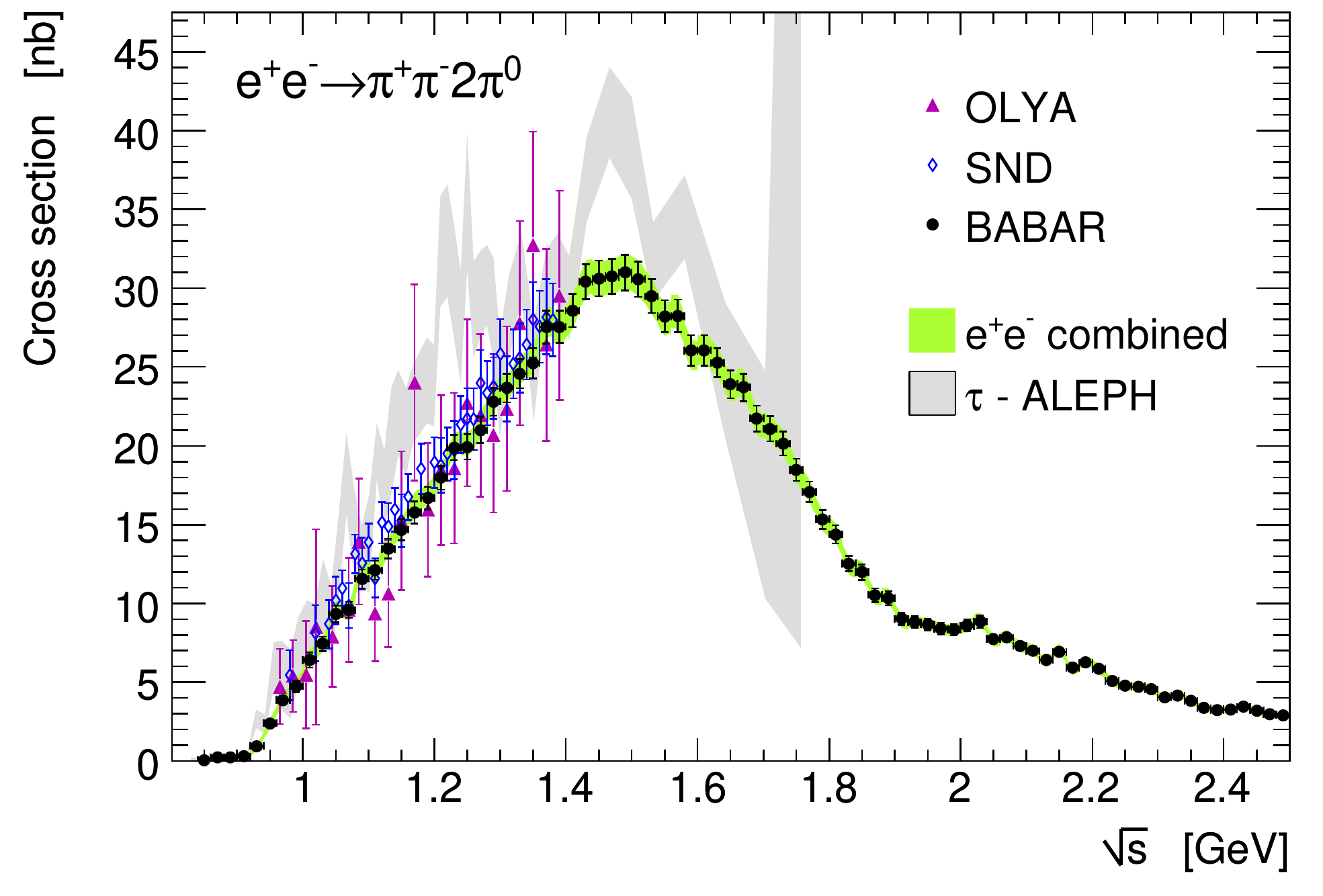}\hspace{\fighspace}
\includegraphics[width=\figsize]{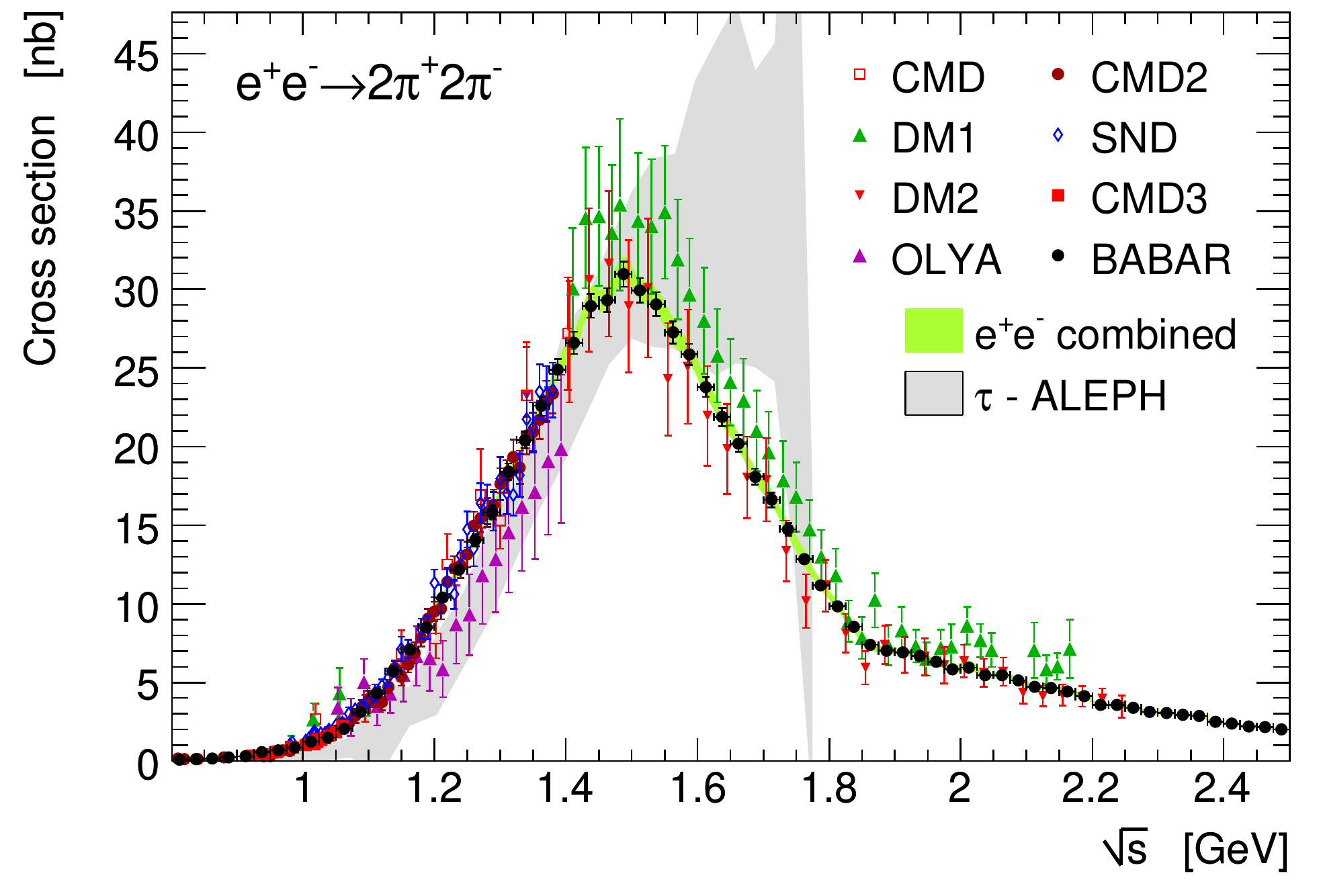}
\end{center}
\vspace{-0.2cm}
\caption[.]{ Bare cross sections for $e^+e^-\to \pi^+ \pi^- 2\pi^0$ (left)
                 and $e^+e^-\to 2\pi^+ 2\pi^- $ (right).
                 The error bars of the data points include statistical and systematic 
                 uncertainties added in quadrature. The green bands show
                 the HVPTools combinations within  $1\,\sigma$ uncertainties. 
                 The cross-section predictions within  $1\,\sigma$  uncertainties
                 derived from ALEPH $\tau$ four-pion spectral functions 
                 are indicated by the light grey bands.
}
\label{fig:4pi}
\vspace{0.4cm}
\begin{center}
\includegraphics[width=\figsize]{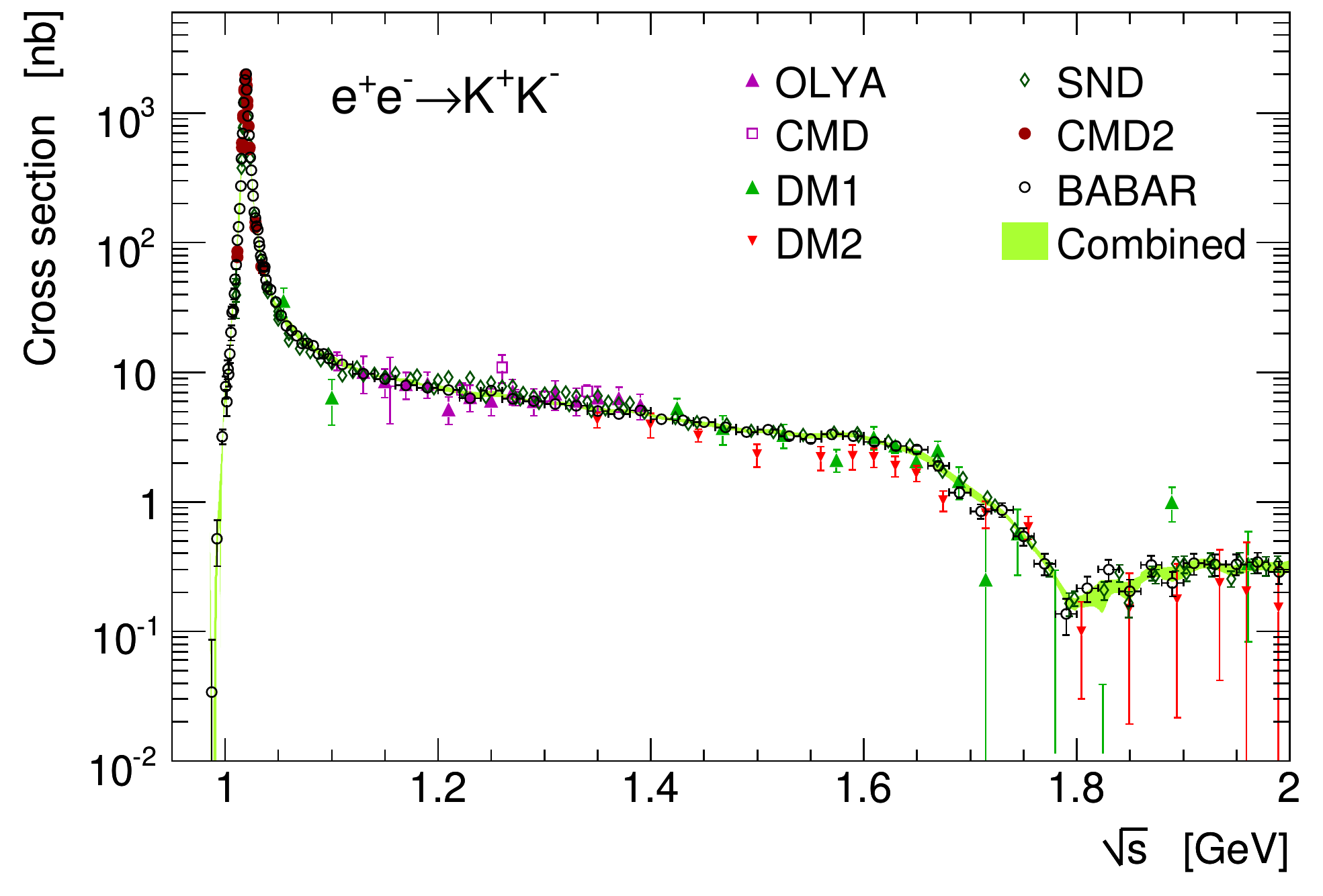} \hspace{\fighspace}
\includegraphics[width=\figsize]{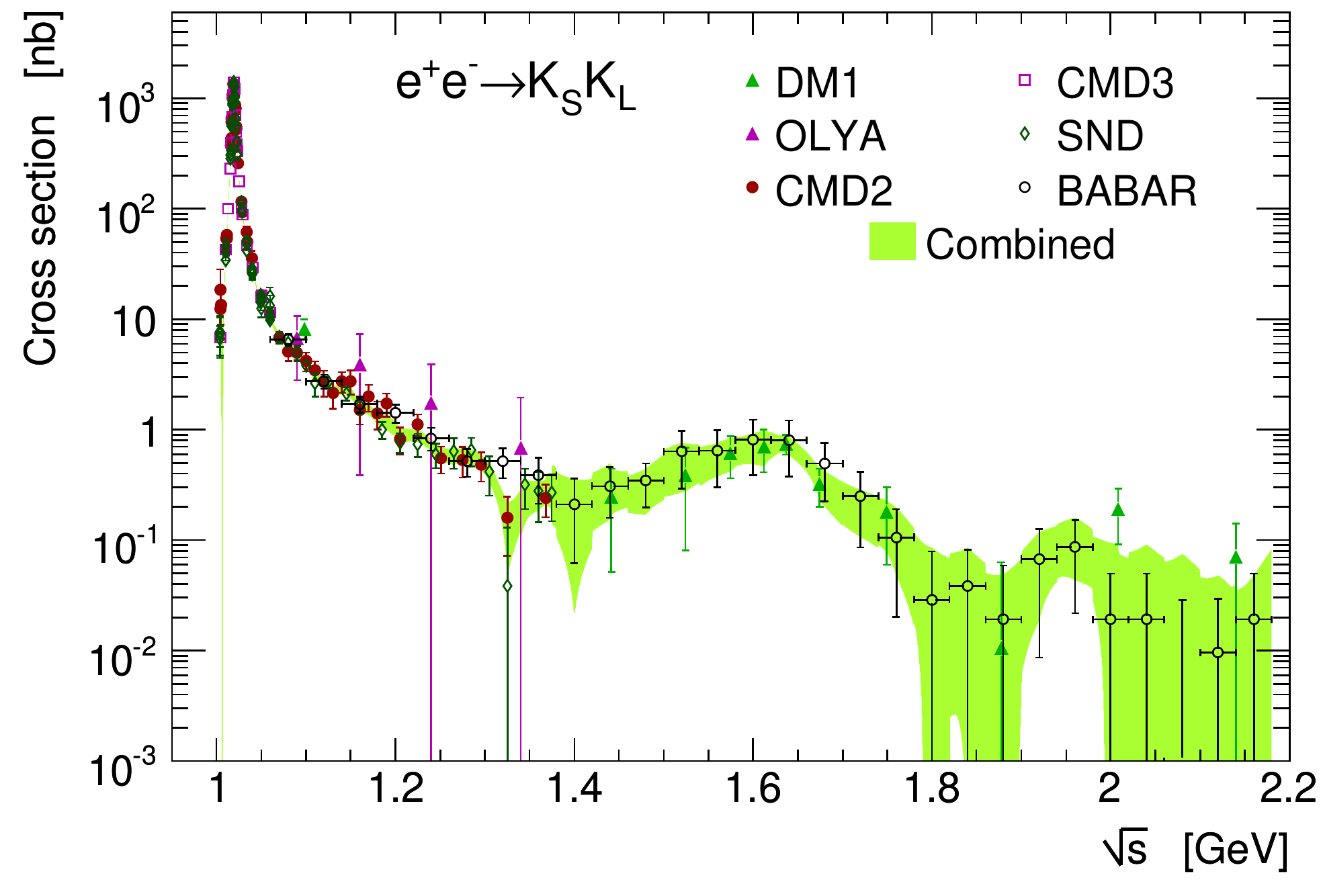}
\end{center}
  \vspace{-0.2cm}
  \caption{
              Bare cross sections for $e^+e^-\to K^+ K^- $ (left) 
              and $e^+e^-\to \KS \KL $  (right). See text for a description of the data used.}
  \label{fig:kk}
\end{figure*}

\begin{figure*}[t]
\begin{center}
 \newcommand\kkpifigsize{0.325\textwidth} 
 \includegraphics[width=\kkpifigsize]{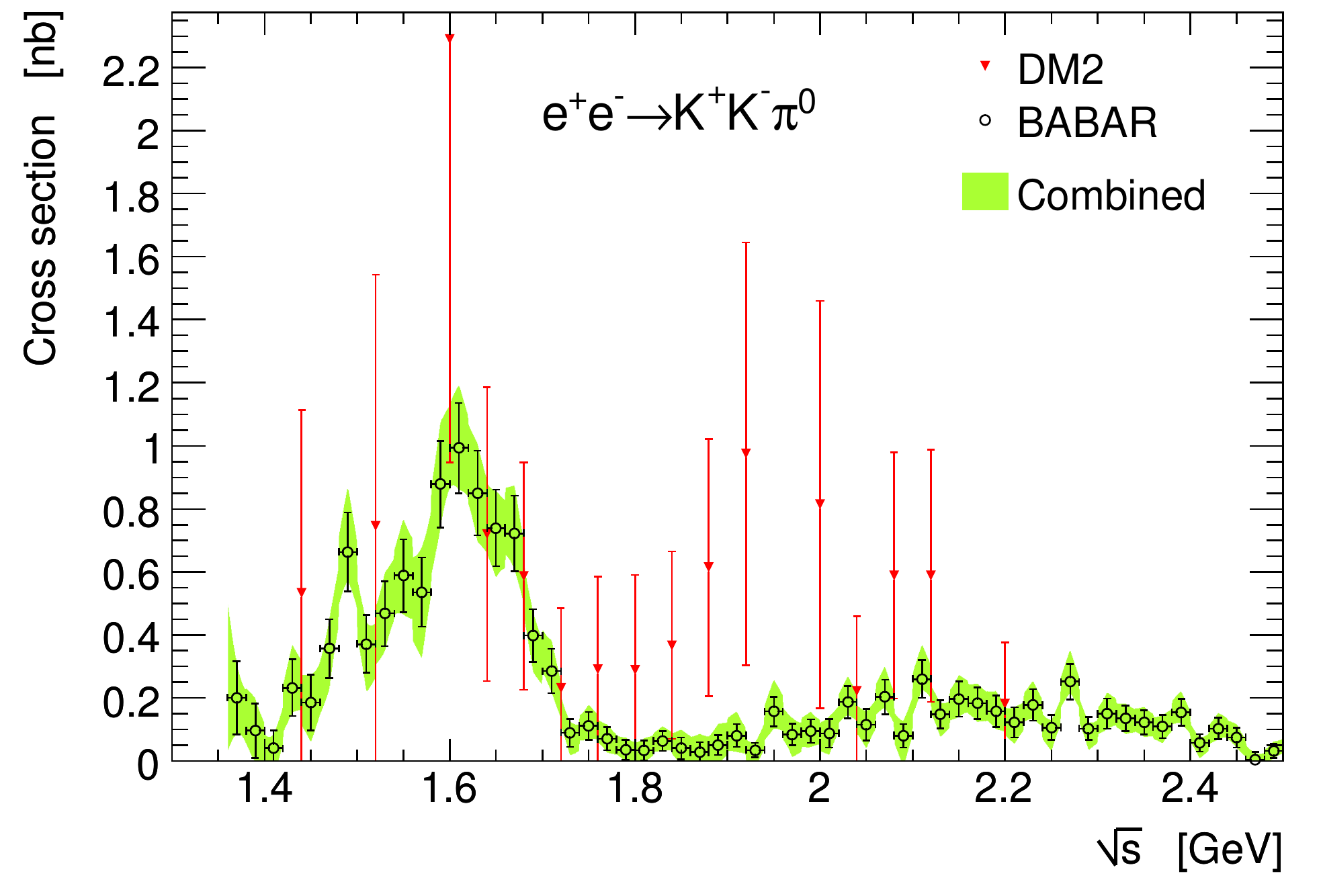}
 \includegraphics[width=\kkpifigsize]{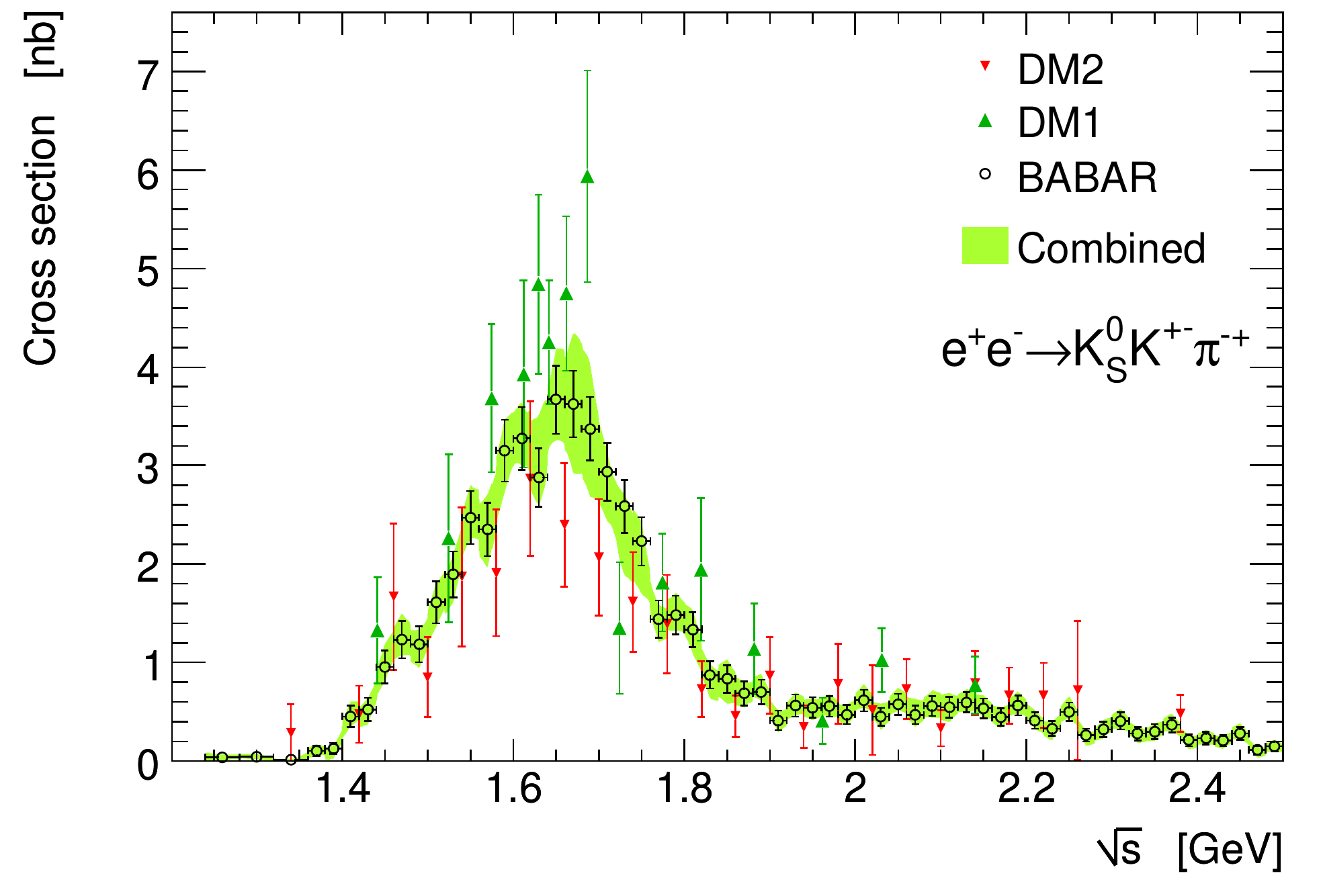}
 \includegraphics[width=\kkpifigsize]{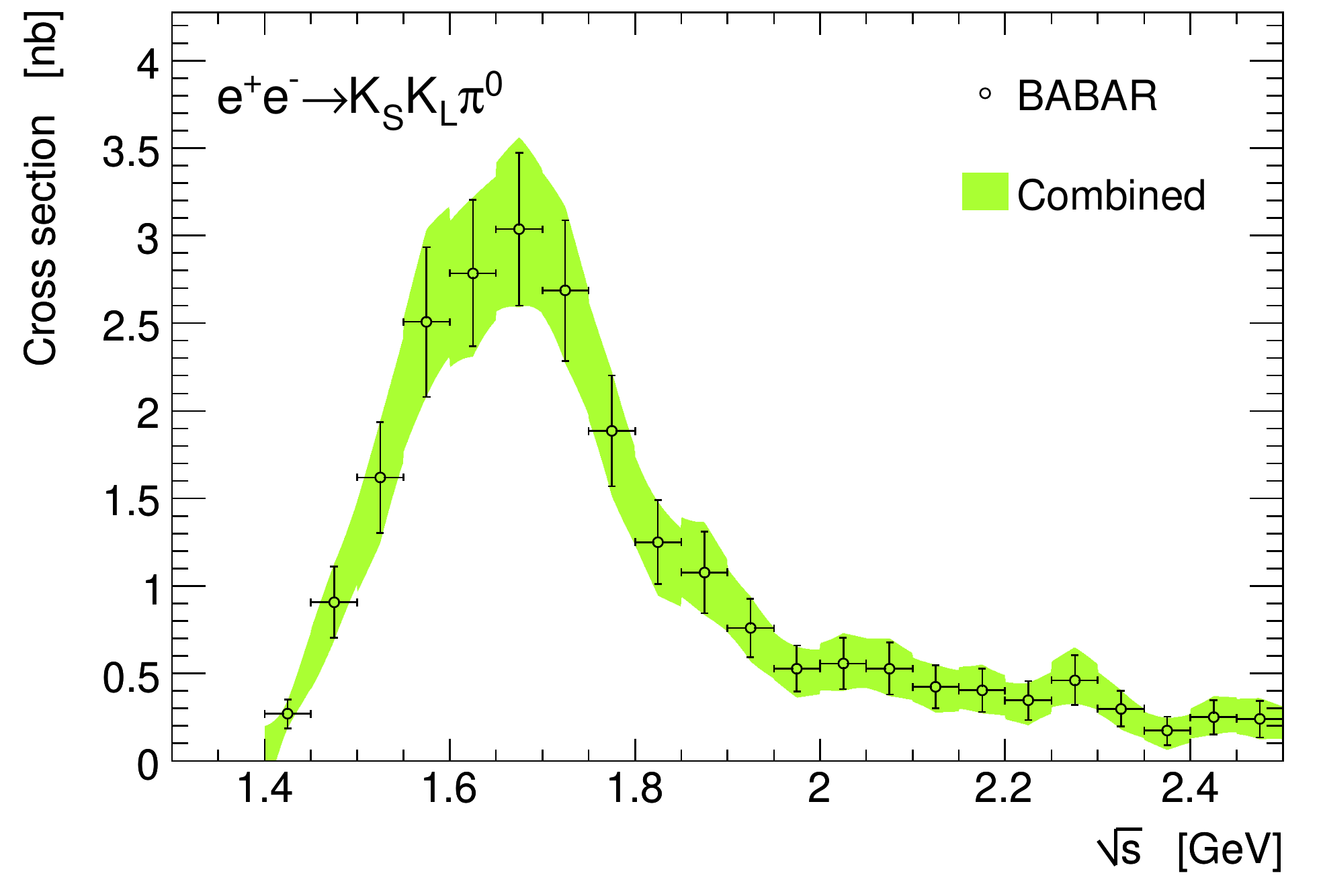} \\
\vspace{0.2cm}
\includegraphics[width=\kkpifigsize]{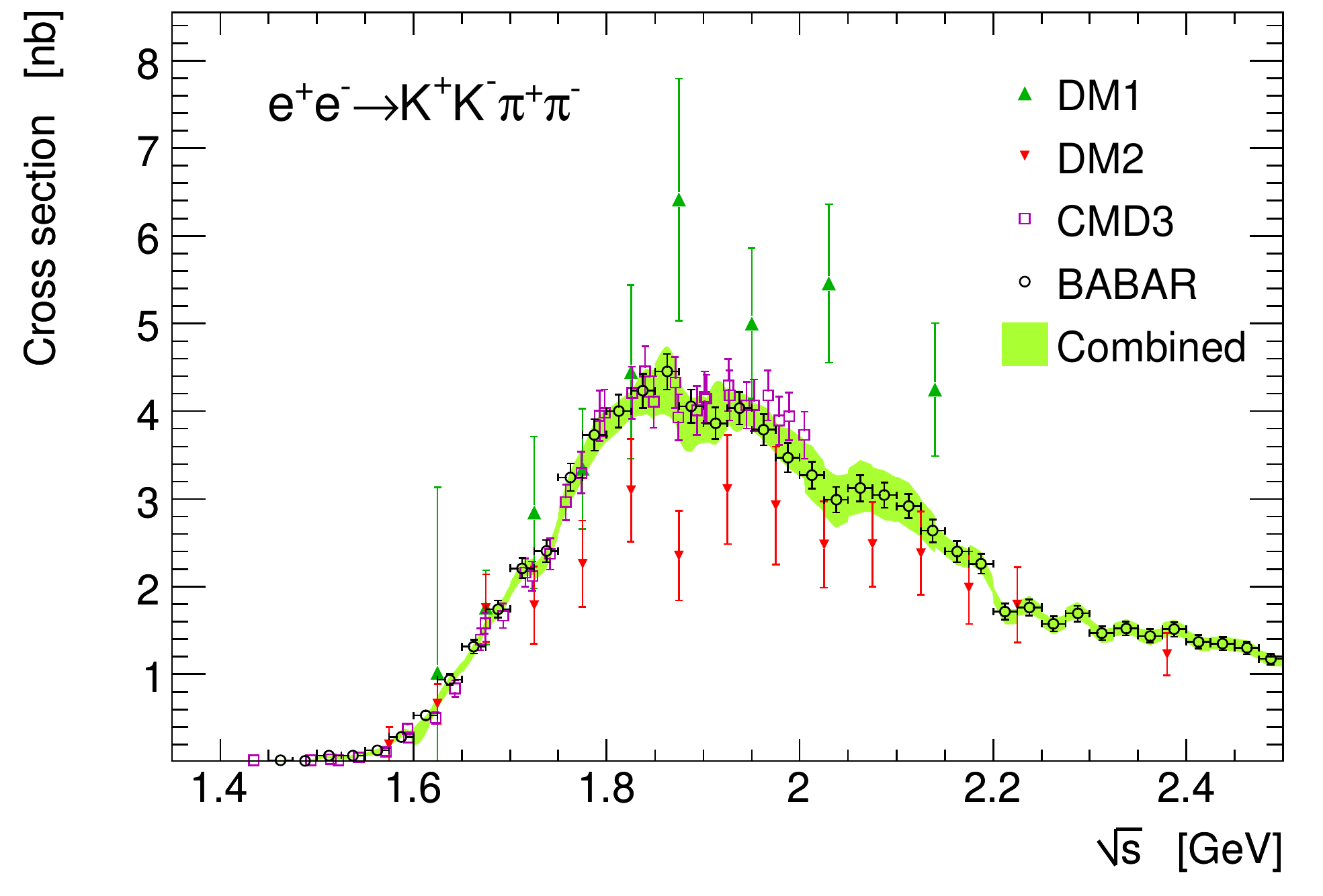}
\includegraphics[width=\kkpifigsize]{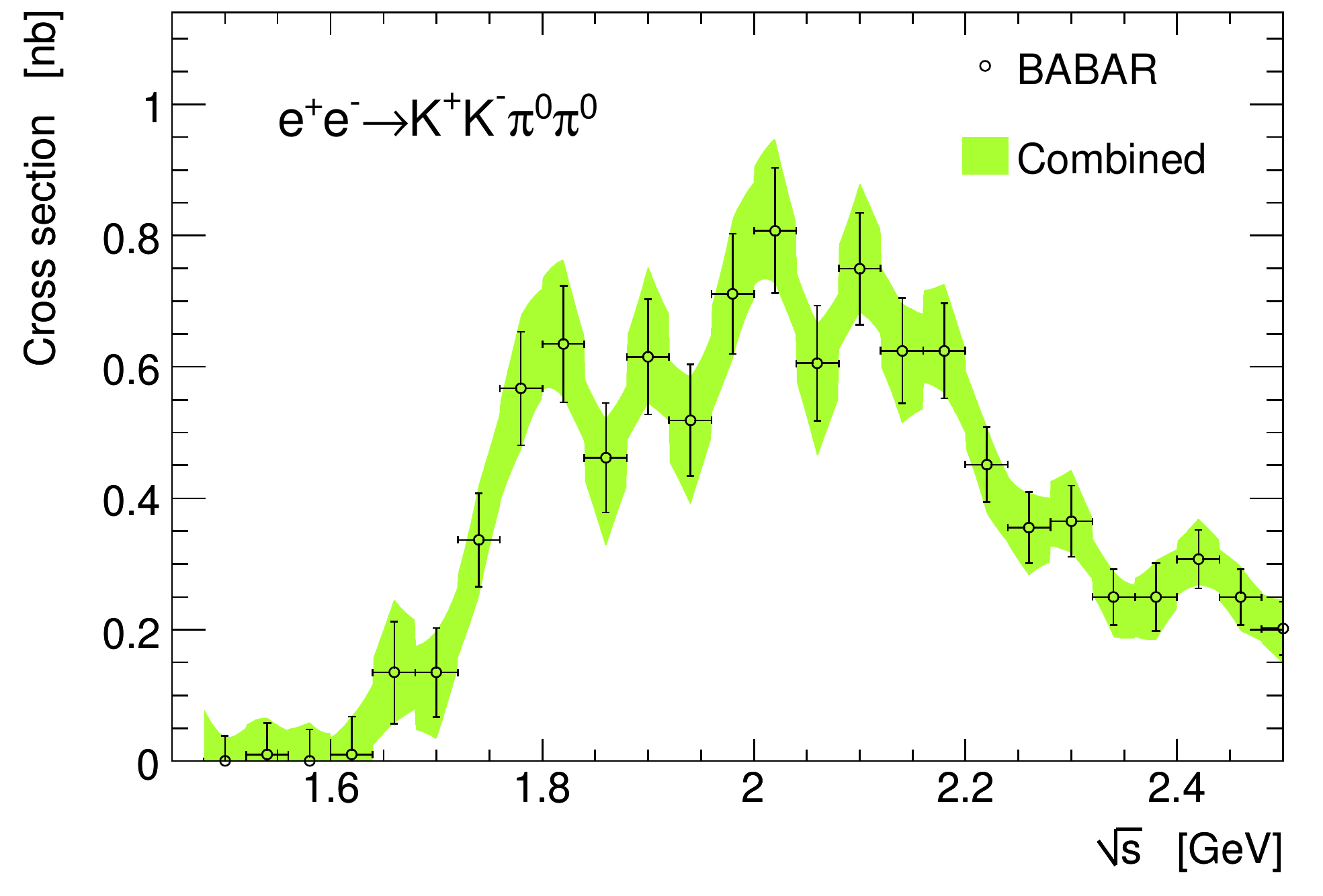}
\includegraphics[width=\kkpifigsize]{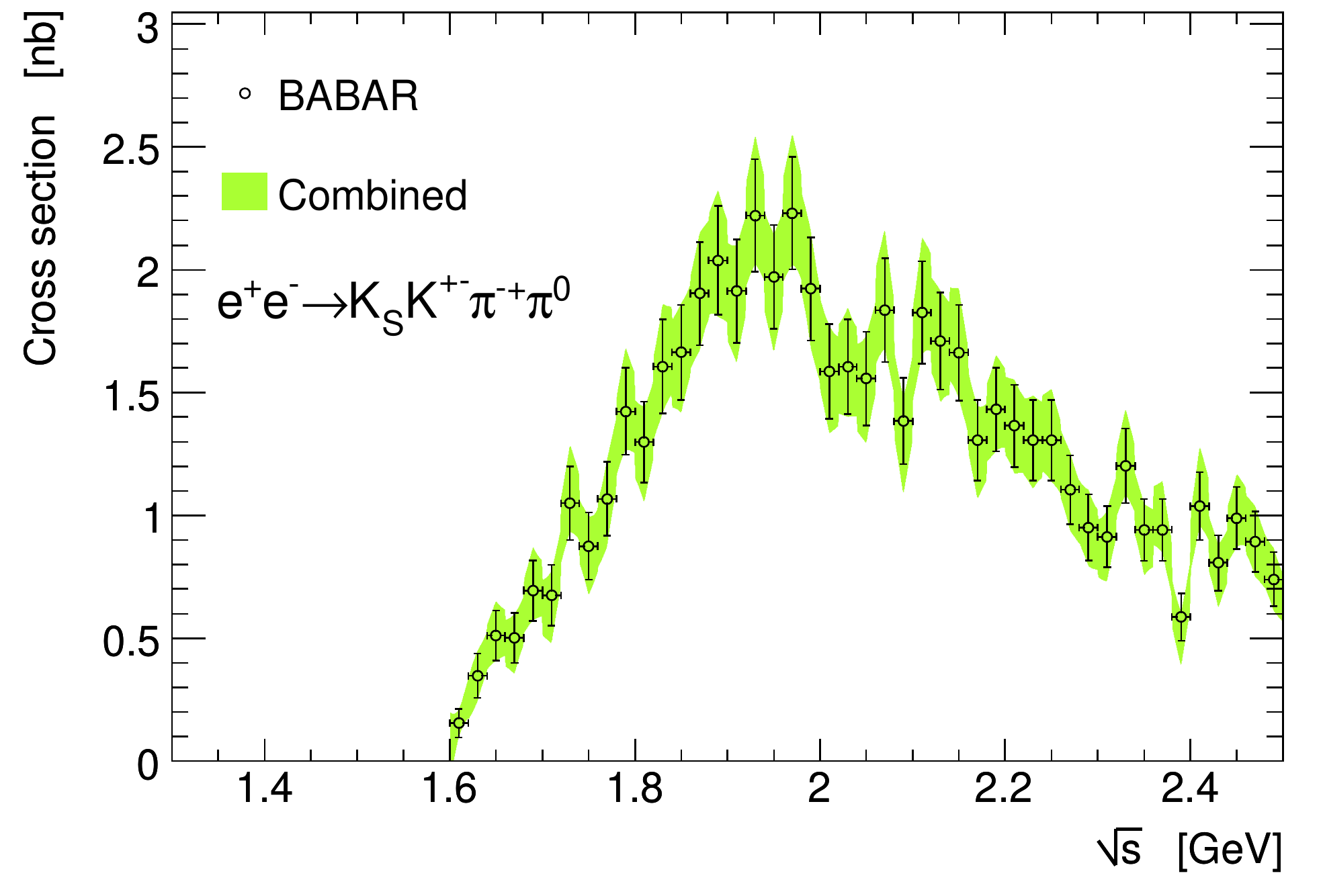}\\
\vspace{0.2cm}
\includegraphics[width=\kkpifigsize]{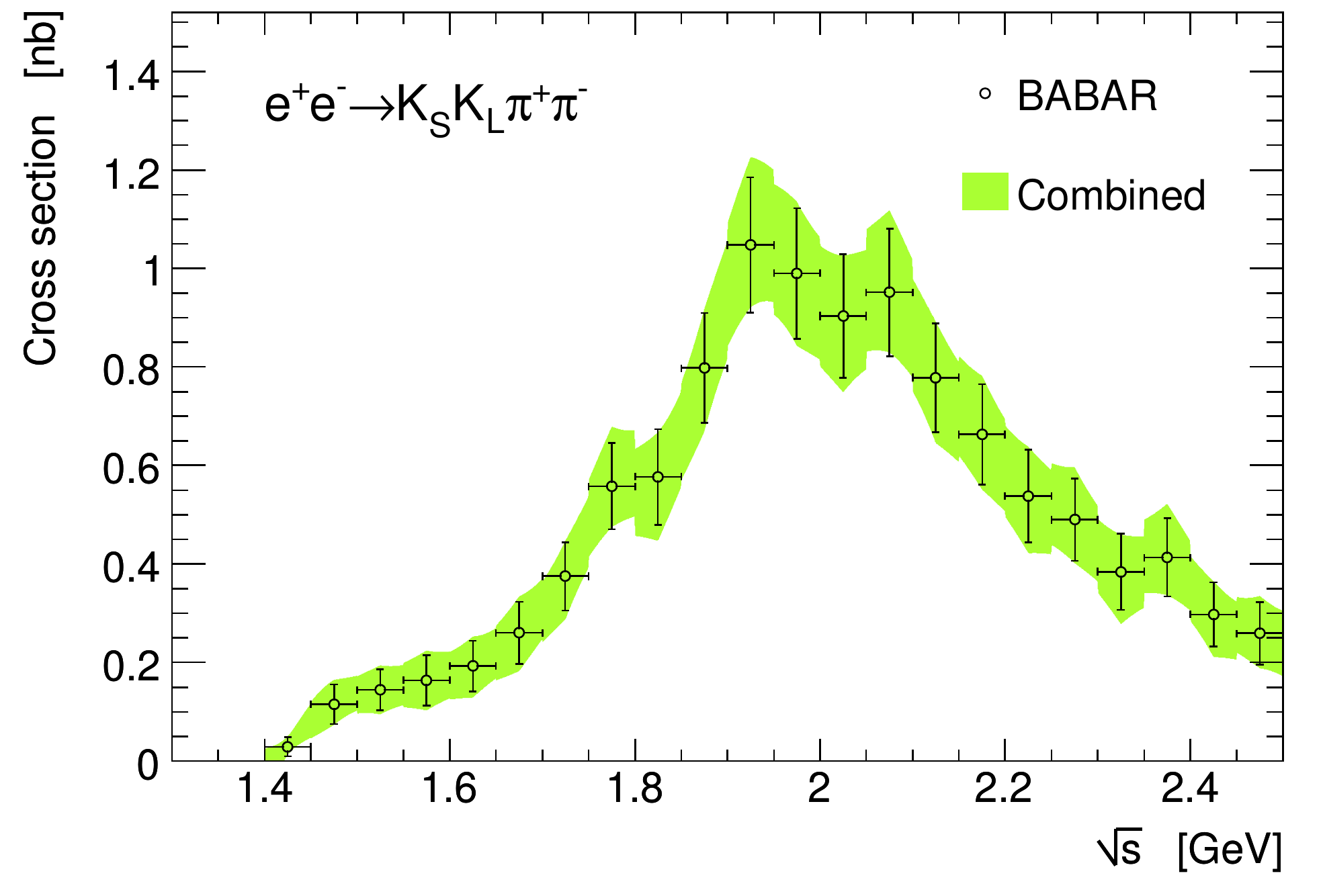}
\includegraphics[width=\kkpifigsize]{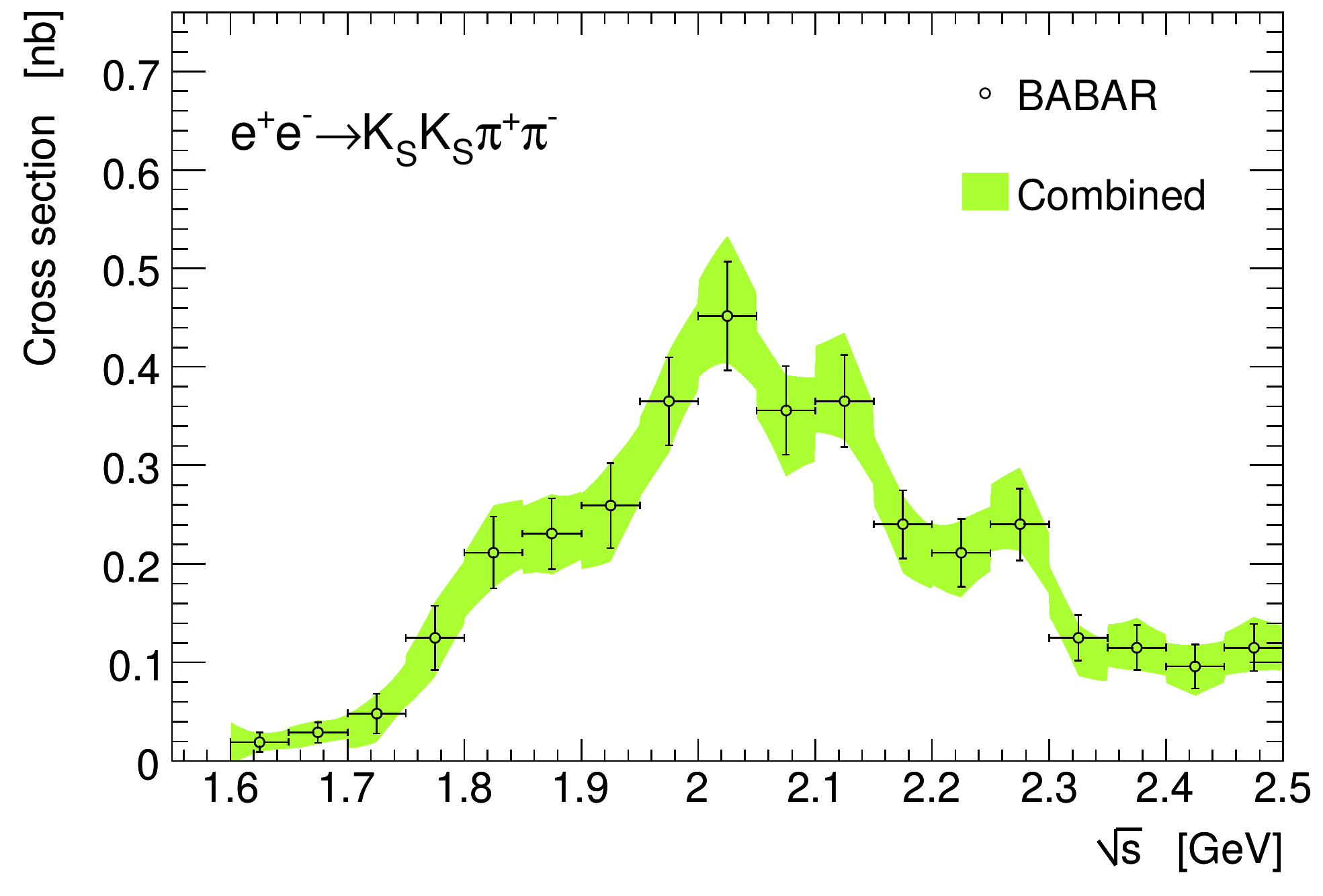}
\includegraphics[width=\kkpifigsize]{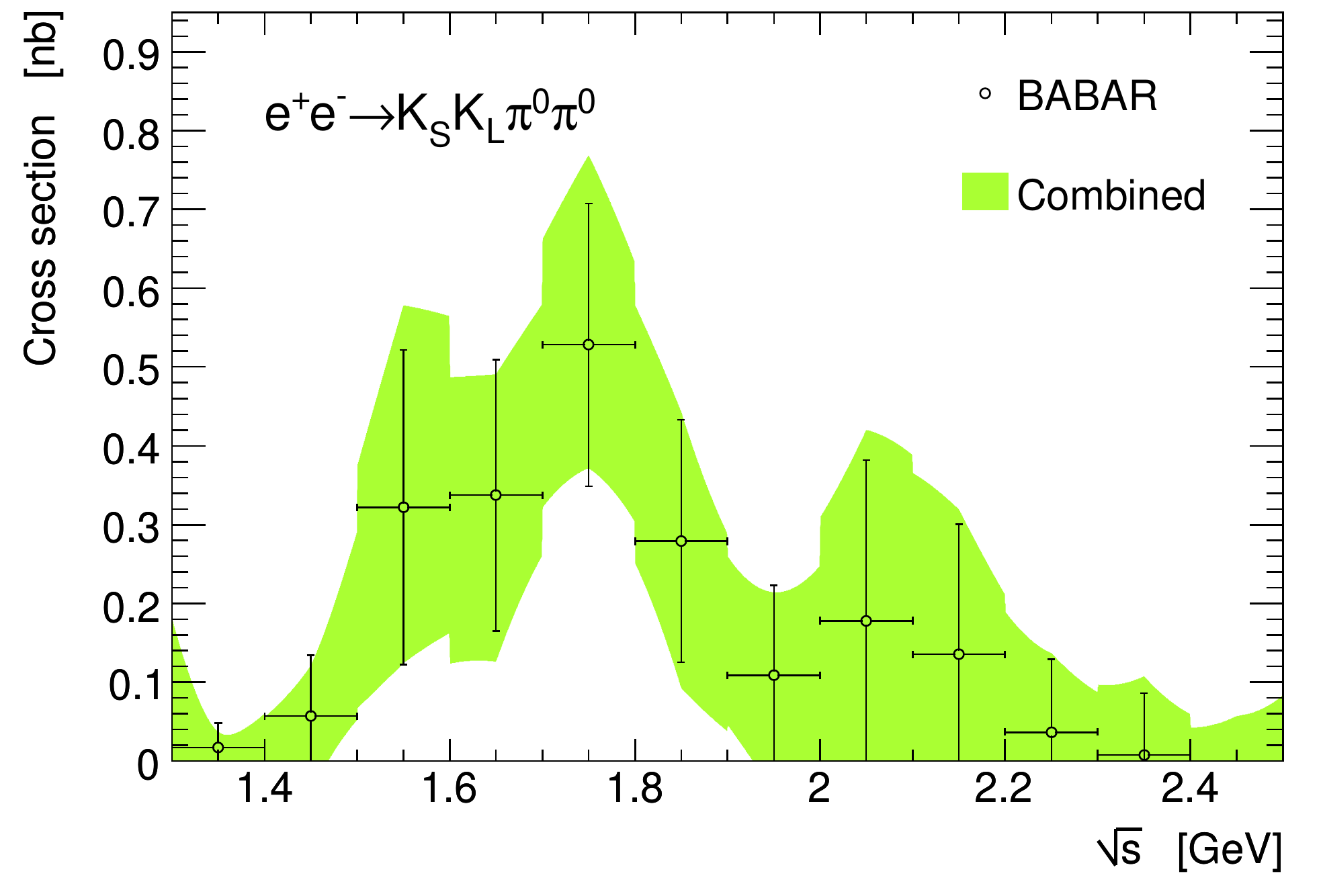}
\end{center}
\vspace{-0.1cm}
\caption[.]{ 
            Bare cross-section data for $K\Kbar\pi$ (top row) and $K\Kbar\pi\pi$ 
            (middle and bottom rows) final states. See text for references.
            The error bars of the data points include statistical and systematic 
            uncertainties added in quadrature. The green bands show the HVPTools 
            interpolations within $1\,\sigma$ uncertainties. Because the integral of the 
            interpolation within a bin of a given measurement is rescaled
            to equal the bin content (recall that the BABAR cross-section measurements are 
            obtained from unfolded histograms) the interpolated cross section can appear
            slightly shifted with respect to the measurement in cases of local 
            shape variations. 
}
  \label{kkbarpi}
\end{figure*}

\begin{figure}[t]
\begin{center}
\includegraphics[width=\figsize]{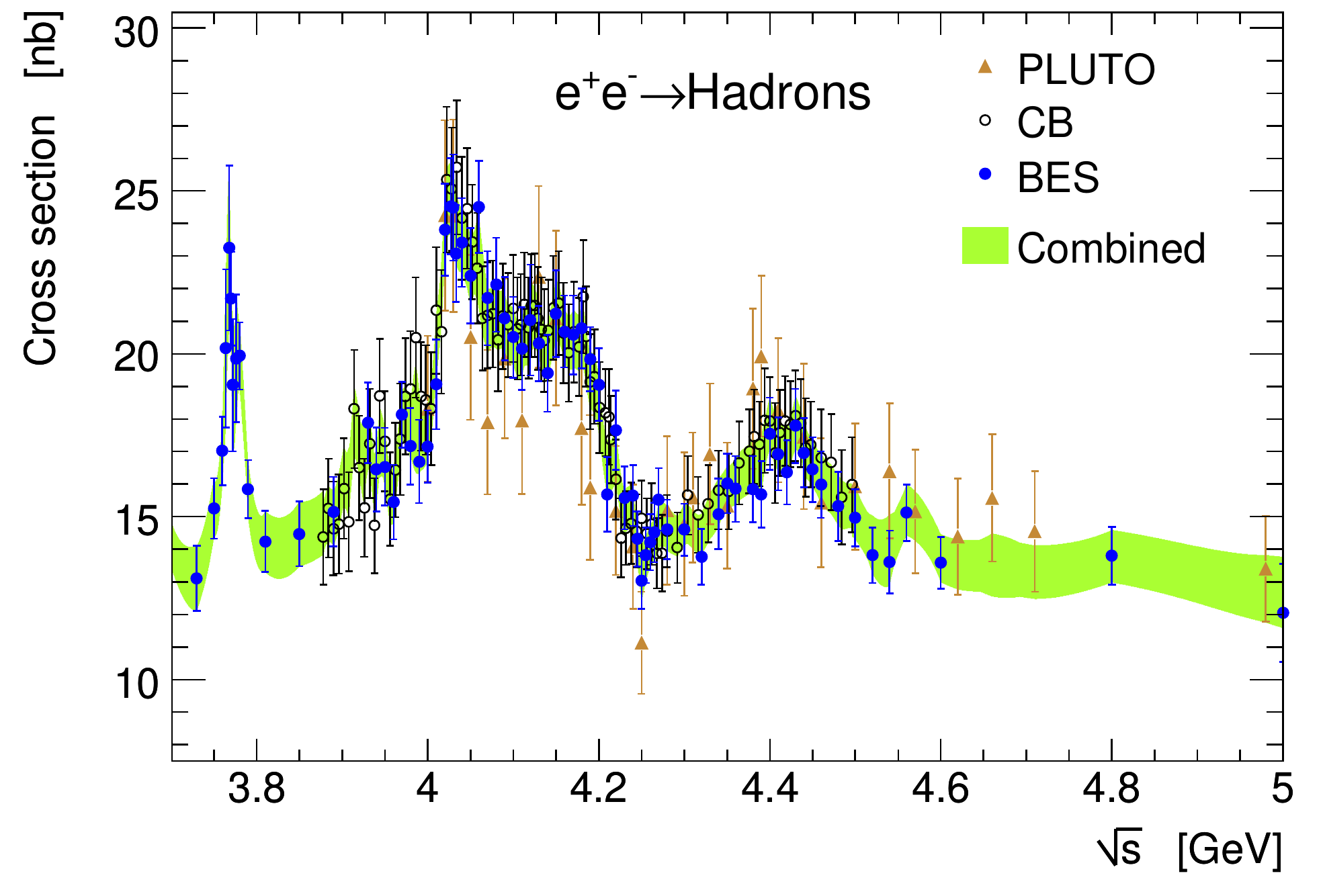}
\end{center}
\vspace{-0.2cm}
\caption[.]{ 
            Inclusive bare hadronic cross section versus centre-of-mass energy above the 
            $D\Dbar$ threshold. 
            The error bars of the data points include statistical and systematic 
            uncertainties added in quadrature. The green band shows
            the HVPTools combination within its $1\,\sigma$ uncertainty. 
}
\label{fig:xseccharm}
\end{figure}

\subsection{~The dominant \sansmath$\mathbf{\boldsymbol{\pi^+\pi^-}}$ channel}
\label{sec:pipi}

The $\pi^+\pi^-$ channel dominates both the HVP contribution to $\amu$ and its 
uncertainty. Recent experiments are generally limited by systematic uncertainties. 
The main contributors are  
BABAR~\cite{babarpipi1,babarpipi2} 
(relative systematic uncertainty of 0.5\% per measurement), 
KLOE-2008~\cite{kloe08} (0.8\%), 
KLOE-2010~\cite{kloe10} (1.4\%), 
CMD2~\cite{cmd203,cmd2new} (0.8\%),
and SND~\cite{snd2pi} (1.5\%),
For this update we newly included 
KLOE-2012~\cite{kloe12} (0.8\%) and the recent 
BESIII-2015~\cite{bes2015} (0.9\%). Only BABAR covers the full \pipi
mass range with high precision. 

The three KLOE measurements exhibit statistical correlations due to the common two-pion events used in the 2008 and 2012 results (the 2012 analysis uses the pion over muon pair cross-section ratio), and systematic correlations from common uncertainty sources.\footnote{Correlations due to systematic uncertainties among the three KLOE datasets are accounted for by matching one-by-one  the uncertainties impacting the  measurements. For example, the uncertainties due to the luminosity, radiator function and vacuum polarisation are taken to be correlated among the first two measurements, while they are found negligible or not present in the 2012 measurement. The final state radiation uncertainty as well as the trigger, tracking, acceptance and background  uncertainties for the $\pi\pi$ data are taken to be correlated among the three measurements, while the background uncertainty for the $\mu\mu$ data only impacts the 2012 measurement. We will replace this approximate treatment by a more accurate one once it is provided by the KLOE Collaboration.} 

Figure~\ref{fig:pipiall} shows the available $\ee\to\pp$ cross-section measurements 
in various panels zooming into different centre-of-mass energies ($\sqrt{s}$). The 
green band indicates the HVPTools combination within its $1\,\sigma$ uncertainty.
The deviations between the combination and the most precise individual measurements are 
plotted in Fig.~\ref{fig:comppipi}. Figure~\ref{fig:weights} (left) shows the local 
combination weight versus $\sqrt{s}$ per experiment. 
The BABAR and KLOE measurements dominate over the entire energy range. 
Owing to the sharp radiator function, the  event yield  increases for KLOE 
towards the $\phi(1020)$ mass, hence outperforming BABAR above $\sim$$0.8\:\gev$.
The group of experiments labelled ``Other exp'' in Fig.~\ref{fig:weights} corresponds 
to older data with incomplete radiative corrections. Their weights are small throughout 
the entire energy domain. The computation of the dispersion integral over the full 
\pp spectrum requires to extend the available data to the region between threshold 
and $0.3\:\gev$, for which we use a fit as described in Ref.~\cite{g209}.

A tension between the BABAR and KLOE measurements is observed at and above the 
$\rho(770)$ peak region (cf. Fig.~\ref{fig:comppipi}), while the other measurements 
are consistent with both. We stress the importance of locally assessing the compatibility 
of the (correlated) cross-section measurements, rather than comparing integrated values 
where discrepancies could cancel or be diluted. The local uncertainty rescaling we apply
(cf. right-hand plot of Fig.~\ref{fig:weights}) increases the combined \amuhadLO 
uncertainty by 15\% in the \pp channel.

In spite of this problem, progress in the evaluation of the \pp contribution to \amuhadLO
has been steady during the last decade. While the central value stayed within 
quoted uncertainties, the  uncertainty dropped from\footnote{Unless  
specified, these and all other \amu related values throughout this paper are given 
in units of $10^{-10}$.} 
5.9 in 2003 to 2.8 in 2011, and now amounts to 2.6. The updated
contribution from threshold to 1.8$\;$GeV is $507.1 \pm 1.1 \pm 2.2 \pm 0.8$, 
where the first uncertainty is statistical, and the second and third stand for 
systematic uncertainties that are, respectively,
uncorrelated and correlated with other channels. The correlation originates mainly 
from uncertainties in the luminosity and in radiative corrections, most notably the 
vacuum polarisation correction applied to the measured cross sections.

Our  \amuhadLO(\pp) estimate using $\tau^-\to\pipiz\nut$  data from ALEPH, OPAL, 
CLEO, and Belle, $516.2 \pm 2.9 \pm 2.2$~\cite{new-aleph-tau}, 
where the first uncertainty is experimental 
and the second due to isospin-breaking corrections, is $2.0\,\sigma$ larger than the 
current \ee-based value. The difference can be reduced by applying off-resonance
$\gamma$--$\rho$ mixing corrections~\cite{jeger-szafron} that come with additional 
uncertainties. Because of the progress in the \ee data, the $\tau$ input is now  
less precise and less reliable due to additional theoretical uncertainties.
While the $\tau$ versus \ee comparison is interesting in its own right, we 
do no longer consider the $\tau$ data for the HVP evaluation. 

\subsection{~The $\pip\pim\piz$ channel}

Following the treatment described in Ref.~\cite{dhmz2011} the contributions 
from $\omega(782)$ and $\phi(1020)$ decaying to three pions are directly evaluated
from the $\ee\to\pp\piz$ cross-section 
measurements (cf. Fig.~\ref{fig:xsec3pi}). Other resonant decays are included 
in the corresponding $\piz\gamma$, $\eta\gamma$, and $K\Kbar$ spectra, while 
small remaining non-resonant decay modes are considered separately. 

\subsection{~The four-pion channels}
\label{sec:4pi}

Recent results using the full BABAR data on $e^+e^-\to\pi^+\pi^-2\pi^0$ are now 
available~\cite{2pi2pi0-babar}. As with other BABAR measurements using the 
ISR method with the ISR photon measured at large angle, the acceptance for the 
recoiling hadronic system is large so that the resonance substructure, dominated 
by the $\omega \pi^0$, $\rho^0 \pi^0 \pi^0$, and $\rho^+ \rho^-$ final states, 
can be fully identified and accurately modelled with a Monte Carlo generator. 
The systematic uncertainty is 3.1\% below 2.7$\;$GeV, a considerable improvement 
over the value of about 10\% of preliminary results available so far. Data from 
some older experiments, both imprecise and inconsistent, are now discarded. As 
seen in the left hand plot of Fig.~\ref{fig:4pi} the BABAR results lead to a 
substantial precision improvement in this channel. 

The  $\pi^+\pi^-2\pi^0$ HVP contribution to \amuhadLO from threshold to 
1.8$\;$GeV is $18.03 \pm 0.06 \pm 0.48 \pm 0.26$, where the total uncertainty of 
0.55 is reduced by a factor of 2.3 compared to our 2011 result~\cite{dhmz2011}.
We note that the $\tau$-based result $21.0 \pm 1.2 \pm 0.4$ (the second
uncertainty accounts for isospin-symmetry breaking corrections), obtained from a 
combination of $\nu_\tau \pi^{-} \pi^+ \pi^-$ and $\nu_\tau \pi^{-} 3\pi^0$ 
spectral functions measured by ALEPH~\cite{new-aleph-tau}, is $2.2\,\sigma$ larger 
than the \ee value and twice less precise.

New $2\pi^+ 2\pi^-$ cross-section data (cf. right hand plot in Fig.~\ref{fig:4pi})
were published by BABAR in 2012~\cite{4pi-babar} 
using the full available data sample and with a reduced systematic uncertainty (2.4\%) 
compared to previous partial results. New measurements from CMD3 between 0.920 and 
1.060$\;$GeV are also available~\cite{cmd3-4pi}. The resulting combined HVP contribution is 
$13.68 \pm 0.03 \pm 0.27 \pm 0.14$, with a total uncertainty of 0.31 reduced 
by a factor of 1.7 compared to our 2011 result~\cite{dhmz2011}.

For comparison, the ALEPH $\tau$-based prediction of $2\pi^+ 2\pi^-$, 
$12.8 \pm 0.7 \pm 0.4$~\cite{new-aleph-tau}, is consistent, but more than twice less 
precise than the \ee-based one.  The $\tau$-based evaluation of the sum 
of the two four-pion channels, $33.8 \pm 1.5$, benefits from an anticorrelation 
due to the $\nu_\tau \pi^{\pm} 3\pi^0$ contribution in both channels. It is consistent 
with the  \ee-based value of $31.7 \pm 0.6$ within $1.3\,\sigma$. The $\tau$-based 
cross-section predictions are compared to the \ee data in Fig.~\ref{fig:4pi}.

\subsection{~The \sansmath$\mathbf{\boldsymbol{K} \kern 0.2em\overline{\kern -0.2em \boldsymbol{K}}{}}$ channels}
\label{sec:KKbar}

New cross-section measurements 
are available for the $\KS\KL$ channel. The BABAR experiment
detects both $\KS$ and $\KL$ from threshold up to 2.2$\;$GeV~\cite{babar-kskl}, while 
CMD3 counts  $\KS$ in the $\phi(1020)$ resonance region~\cite{cmd3-kskl}.
Consistency is observed between the two experiments as well as with older measurements
from CMD2 and SND. The measured cross sections are shown in the right panel of 
Fig.~\ref{fig:kk}.

The  $\KS\KL$ contribution to \amuhadLO up to 1.8$\;$GeV amounts to
$12.81 \pm 0.06 \pm 0.18 \pm 0.15$ with a total uncertainty of 0.24, which is  
reduced by a factor of 1.6 over that of our 2011 estimate~\cite{dhmz2011}.

Recent measurements from SND~\cite{snd-kpkm} at VEPP-2000 for the $K^+K^-$ channel
agree with BABAR~\cite{babar-kpkm}, while both show a discrepancy with  former 
SND data, obtained at VEPP-2M below 1.4$\;$GeV, that exceeds the quoted systematic
uncertainty. The BABAR and new SND data are displayed in the left hand panel 
of Fig.~\ref{fig:kk}.

Some concern arises with regard to the $e^+e^-\to\phi\to K^+K^-$ cross-section
measurements. The BABAR result has a systematic uncertainty of 0.7\%, but it is 
larger by 5.1\% (9.6\%) than CMD2 (SND) which has a systematic uncertainty of 2.2\% 
(7.1\%). Including the BABAR data the contribution to \amuhadLO
increases from 21.63 to 22.67 with an uncertainty of 0.43. A recent
preliminary result from CMD3~\cite{cmd3-kpkm} shows a very large ($\sim$11\%) 
excess of the cross section over CMD2 and $\sim$5\% over BABAR. The origin of this
large discrepancy is not understood at present~\cite{solodov-private}. It raises doubts 
on the ability to perform a precision measurement of this channel  
with the standard energy-scan method because the detection efficiency of the 
low-momentum $K^\pm$ from  $\phi$ decay  is difficult to model. Owing to the boost 
of the final state, the ISR method is expected to be more reliable for the charged 
kaon detection.

\subsection{~The \sansmath$\mathbf{\boldsymbol{K} \kern 0.2em\overline{\kern -0.2em \boldsymbol{K}}{}} \:+ \:$pions channels}
\label{sec:KKbar-pions}

In previous hadronic vacuum polarisation analyses the available exclusive 
$e^+e^- \to K\Kbar\:+\:$pions data were incomplete. Missing channels were 
constrained based on assumptions about the process dynamics and isospin 
symmetry~\cite{dehz2003,dhmz2011} leading to considerable uncertainty. This 
procedure became unnecessary since the BABAR experiment  produced cross-section 
results for the three channels contributing to the final state $K\Kbar\pi$ 
and six channels contributing to $K\Kbar\pi\pi$. 
A key ingredient of the BABAR analyses is the detection of neutral kaons, both 
$\KS$, through the $\pi^+\pi^-$ decay, and $\KL$ interacting in the calorimeters. 

Together with previous measurements of $\KS K^{\pm} \pi^{\mp}$ and $K^+K^- \pi^0$, data 
for $\KS\KL \pi^0$~\cite{babar-ksklpi0}  complete the picture for the $K\Kbar\pi$ channel
(cf. top row panels in Fig.~\ref{kkbarpi}). Because that final state is dominated 
by $K^\star \Kbar$ 
production below 1.8$\;$GeV (with a small contribution from $\phi \pi^0$), it is 
expected that isospin symmetry provides a good approximation. Indeed, the contribution 
from the sum of the measured channels, $2.45 \pm 0.15$, is in agreement with and 
has similar precision as the result $2.39 \pm 0.16$ obtained from the 
$\KS K^{\pm} \pi^{\mp}$ data  only together with isospin symmetry.

Of the six channels contributing to $K\Kbar\pi\pi$ only two, $K^+K^-\pi^+\pi^-$ and 
$K^+K^- 2\pi^0$, had been measured by BABAR in 2011. Constraints from isospin symmetry 
were used to estimate the missing channels~\cite{dhmz2011}, 
but because of the complex dynamics involving $K^*(890) \Kbar  \pi$, $K \Kbar  \rho$ 
and $\phi \pi \pi$ intermediate states, these estimates were 
plagued by substantial uncertainties. Among the remaining channels, 
$\KS\KL\pi^+\pi^-$~\cite{babar-kskl}, $\KS\KS \pi^+\pi^-$~\cite{babar-kskl}, 
$\KS\KL 2\pi^0$~\cite{babar-ksklpi0}, and 
$\KS K^{\pm} \pi^{\pm} \pi^0$~\cite{babar-KsKpipi0} have been measured by BABAR.  
In addition, the previously measured channels $K^+K^-\pi^+\pi^-$ and $K^+K^- 2\pi^0$
have been updated with the full data sample~\cite{babar-kkpipi}.  
New measurements of $K^+K^-\pi^+\pi^-$ became available 
from CMD3~\cite{cmd3-kkpipi} and are in  agreement with the BABAR data. 
Except for $\KL\KL \pi^+\pi^-$, which can be safely estimated using CP symmetry, 
all cross sections have now been measured. 

The precision in the inclusive contribution to \amuhadLO from all $K\Kbar\pi\pi$ 
final states improved from previously 0.39, dominated by the uncertainty in 
the estimates from isospin symmetry~\cite{dhmz2011}, to presently 0.05 
(cf. Table~\ref{tab:results} on page~\pageref{tab:results}).

\subsection{~Other channels}
\label{sec:others}

Data on many processes with smaller cross sections, mainly from VEPP-2000, have become  
available and are included in the HVPTools database. This is the case for $3\pi^+3\pi^-$ from 
CMD3~\cite{cmd3-6pi}, $\pi^0\gamma$~\cite{snd-pi0g}, $\eta\gamma$~\cite{snd-etag}, 
$\eta\pi^+\pi^-$~\cite{snd-etapipi}, and $\omega\pi^0$~\cite{snd-omegapi} from SND,
$\eta\omega$ from SND~\cite{snd-etaomega} and CMD3~\cite{cmd3-etaomega}, 
nonresonant $\eta\pi^+\pi^-\pi^0$ from CMD3~\cite{cmd3-etaomega},
 $\eta\pi^+\pi^-$ from BABAR~\cite{babar-etapipi},
which extend and improve older measurements in these channels. 
Except for the $\eta\omega$ cross section above 1.6$\;$GeV,
results using the ISR technique at BABAR and the scan method at VEPP-2000 are in 
agreement notwithstanding 
their different systematic uncertainties. Above 1.8$\;$GeV the production of
$p\pbar$ measured by BABAR~\cite{babar-ppbar} and CMD3~\cite{cmd3-ppbar}, $n\nbar$
by ADONE~\cite{adone-nnbar} and SND~\cite{snd-nnbar}, and $\eta\omega\pi^0$ by
SND~\cite{snd-etaomegapi} are included. 

Figure~\ref{fig:xseccharm} shows the available measurements and their combination 
of the charm resonance region above the opening of the $D\Dbar$ channel.  
The individual datasets agree within 
uncertainties. While Crystal Ball~\cite{cbR} and BES~\cite{besR}
published bare inclusive cross-section results, PLUTO applied only radiative 
corrections~\cite{plutoradcorr}
following the formalism of Ref.~\cite{bonnmartin}, which does not include HVP.
Following similar previous cases~\cite{dehz02}, we have applied this 
correction and assigned a 50\% systematic uncertainty to it.

\subsection{~Estimated missing channels}
\label{sec:missing}

Even with the large number of exclusive cross-section measurements 
available below 2$\;$GeV, covering 
particle multiplicities up to six hadrons including $\pi^0$ and $\eta$ mesons, a few channels 
with more than two neutral pions are still unmeasured and their contributions must be 
estimated using isospin symmetry. The treatment of the channels 
$\pi^+\pi^-3\pi^0$, $\pi^+\pi^-4\pi^0$, and $\eta \pi^+\pi^- 2\pi^0$   follows our 
previous approach detailed in Ref.~\cite{dhmz2011}. 

Whereas the $e^+e^-\to \eta\phi$ cross-section data were already included, 
the previously neglected 
smaller contribution from $e^+e^-\to \eta (K^+K^-)_{\mbox{\scriptsize non-}\phi}$ 
where the $K^+K^-$ does not originate from resonant $\phi$ decay is now taken into account 
following a BABAR measurement~\cite{babar-KsKpi}. Its unmeasured counterpart 
$e^+e^-\to \eta (K_SK_L)_{\mbox{\scriptsize non-}\phi}$ can be crudely estimated 
to equal the corresponding  $K^+K^-$ contribution with a 100\% systematic uncertainty. 
This estimate is consistent with the upper limit that can be deduced using BABAR's 
$\KS\KL\eta$ measurement~\cite{babar-ksklpi0}.


Altogether the contribution from missing channels to \amuhadLO up to 1.8$\;$GeV is 
estimated to be $0.46 \pm 0.12$, corresponding to a fraction of only $(0.09 \pm 0.02)$\% 
of the full HVP contribution in this range. The corresponding fraction in our 2011 
analysis was much larger, $(0.69 \pm 0.07)$\%, illustrating the experimental progress made.

\section{~Compilation and results}
\label{sec:Results}

\begin{table*}[p]
\newcommand{\gam}{\ensuremath{\gamma}\xspace}
\setlength{\tabcolsep}{0.0pc}
\begin{tabularx}{\textwidth}{@{\extracolsep{\fill}}lrr} 
\hline\noalign{\smallskip}
Channel &   \amuhadLO $[10^{-10}]$ & \dahadZ $[10^{-4}]$ \\
\noalign{\smallskip}\hline\noalign{\smallskip}
$\pi^0\gam$                                               &$  4.29 \pm 0.06 \pm 0.04 \pm 0.07$&$  0.35 \pm 0.00 \pm 0.00 \pm 0.01$ \\
$\eta\gam$                                                &$  0.65 \pm 0.02 \pm 0.01 \pm 0.01$&$  0.08 \pm 0.00 \pm 0.00 \pm 0.00$ \\
$\pi^+\pi^-$                                              &$507.14 \pm 1.13 \pm 2.20 \pm 0.75$&$ 34.39 \pm 0.07 \pm 0.15 \pm 0.05$ \\
$\pi^+\pi^-\pi^0$                                         &$ 46.20 \pm 0.40 \pm 1.10 \pm 0.86$&$  4.60 \pm 0.04 \pm 0.11 \pm 0.08$ \\
$2\pi^+2\pi^-$                                            &$ 13.68 \pm 0.03 \pm 0.27 \pm 0.14$&$  3.58 \pm 0.01 \pm 0.07 \pm 0.03$ \\
$\pi^+\pi^-2\pi^0$                                        &$ 18.03 \pm 0.06 \pm 0.48 \pm 0.26$&$  4.45 \pm 0.02 \pm 0.12 \pm 0.07$ \\
$2\pi^+2\pi^-\pi^0$ ($\eta$ excl.)                        &$  0.69 \pm 0.04 \pm 0.06 \pm 0.03$&$  0.21 \pm 0.01 \pm 0.02 \pm 0.01$\\
$\pi^+\pi^-3\pi^0$ ($\eta$ excl.,  isospin)               &$  0.35 \pm 0.02 \pm 0.03 \pm 0.01$&$  0.11 \pm 0.01 \pm 0.01 \pm 0.00$\\
$3\pi^+3\pi^-$                                            &$  0.11 \pm 0.00 \pm 0.01 \pm 0.00$&$  0.04 \pm 0.00 \pm 0.00 \pm 0.00$ \\
$2\pi^+2\pi^-2\pi^0$  ($\eta$ excl.)                      &$  0.72 \pm 0.06 \pm 0.07 \pm 0.14$&$  0.25 \pm 0.02 \pm 0.02 \pm 0.05$\\
$\pi^+\pi^-4\pi^0$ ($\eta$ excl.,  isospin)               &$  0.11 \pm 0.01 \pm 0.11 \pm 0.00$&$  0.04 \pm 0.00 \pm 0.04 \pm 0.00$\\
$\eta\pi^+\pi^-$                                          &$  1.18 \pm 0.03 \pm 0.06 \pm 0.02$&$  0.34 \pm 0.01 \pm 0.02 \pm 0.01$ \\
$\eta\omega$                                              &$  0.32 \pm 0.02 \pm 0.02 \pm 0.01$&$  0.10 \pm 0.01 \pm 0.01 \pm 0.00$\\
$\eta \pi^+\pi^-\pi^0$ (non-$\omega,\phi$)                &$  0.39 \pm 0.03 \pm 0.11 \pm 0.03$&$  0.13 \pm 0.01 \pm 0.04 \pm 0.01$\\
$\eta 2\pi^+2\pi^-$                                       &$  0.03 \pm 0.01 \pm 0.00 \pm 0.00$&$  0.01 \pm 0.00 \pm 0.00 \pm 0.00$ \\
$\eta\pi^+\pi^-2\pi^0$                                    &$  0.03 \pm 0.01 \pm 0.01 \pm 0.00$&$  0.01 \pm 0.00 \pm 0.00 \pm 0.00$ \\
$\omega\pi^0~(\omega\to\pi^0\gam)$                        &$  0.94 \pm 0.01 \pm 0.02 \pm 0.02$&$  0.20 \pm 0.00 \pm 0.00 \pm 0.00$\\
$\omega(\pi\pi)^0~(\omega\to\pi^0\gam)$                   &$  0.08 \pm 0.00 \pm 0.01 \pm 0.00$&$  0.03 \pm 0.00 \pm 0.00 \pm 0.00$\\
$\omega$ (non-$3\pi,\pi\gam,\eta\gam$)                    &$  0.36 \pm 0.00 \pm 0.01 \pm 0.00$&$  0.03 \pm 0.00 \pm 0.00 \pm 0.00$\\
$K^+K^-$                                                  &$ 22.81 \pm 0.24 \pm 0.28 \pm 0.17$&$  3.31 \pm 0.03 \pm 0.04 \pm 0.03$\\
$K_SK_L$                                                  &$ 12.82 \pm 0.06 \pm 0.18 \pm 0.15$&$  1.74 \pm 0.01 \pm 0.03 \pm 0.02$\\
$\phi$ (non-$K\Kbar,3\pi,\pi\gam,\eta\gam$)               &$  0.05 \pm 0.00 \pm 0.00 \pm 0.00$&$  0.01 \pm 0.00 \pm 0.00 \pm 0.00$\\
$K\Kbar\pi$                                               &$  2.45 \pm 0.06 \pm 0.12 \pm 0.07$&$  0.78 \pm0.02  \pm 0.04 \pm 0.02$\\
$K\Kbar2\pi$                                              &$  0.85 \pm 0.02 \pm 0.05 \pm 0.01$&$  0.30 \pm0.01  \pm 0.02 \pm 0.00$\\
$K\Kbar3\pi$~(estimate)                                   &$ -0.03 \pm 0.01 \pm 0.02 \pm 0.00$&$ -0.01 \pm 0.00 \pm 0.01 \pm 0.00$\\
$\eta\phi$                                                &$  0.36 \pm 0.02 \pm 0.02 \pm 0.01$&$  0.13 \pm 0.01 \pm 0.01 \pm 0.00$\\
$\eta K\Kbar$ (non-$\phi$)                                &$  0.01 \pm 0.01 \pm 0.01 \pm 0.00$&$  0.00 \pm 0.00 \pm 0.01 \pm 0.00$\\
$\omega K\Kbar~(\omega\to\pi^0\gam)$                      &$  0.01 \pm 0.00 \pm 0.00 \pm 0.00$&$  0.00 \pm 0.00 \pm 0.00 \pm 0.00$\\
$\omega\eta\pi^0$                                         &$  0.06 \pm 0.04 \pm 0.00 \pm 0.00$&$  0.02 \pm 0.02 \pm 0.00 \pm 0.00$\\
\noalign{\smallskip}\hline\noalign{\smallskip}        
$J/\psi$  (BW integral)                                   &$  6.28 \pm 0.07$&$  7.09 \pm 0.08$  \\
$\psi(2S)$  (BW integral)                                 &$  1.57 \pm 0.03$&$  2.50 \pm 0.04$  \\
\noalign{\smallskip}\hline\noalign{\smallskip}
$R_{\rm data}~~[3.7$--$5.0\:\gev]$                        &$  7.29 \pm 0.05 \pm 0.30 \pm 0.00$&$ 15.79 \pm 0.12 \pm 0.66 \pm 0.00$\\
\noalign{\smallskip}\hline\noalign{\smallskip}
$R_{\rm QCD}~~[1.8$--$3.7\:\gev]_{uds}$                   &$ 33.45 \pm 0.28 \pm 0.59_{\rm dual}$&$ 24.27 \pm 0.18 \pm 0.26_{\rm dual}$ \\
$R_{\rm QCD}~~[5.0$--$9.3\:\gev]_{udsc}$                  &$  6.86 \pm 0.04$&$ 34.89 \pm 0.17$\\
$R_{\rm QCD}~~[9.3$--$12.0\:\gev]_{udscb}$                &$  1.21 \pm 0.01$&$ 15.56 \pm 0.04$\\
$R_{\rm QCD}~~[12.0$--$40.0\:\gev]_{udscb}$               &$  1.64 \pm 0.00$&$ 77.94 \pm 0.12$\\
$R_{\rm QCD}~~[>40.0\:\gev]_{udscb}$                      &$  0.16 \pm 0.00$&$ 42.70 \pm 0.06$\\
$R_{\rm QCD}~~[>40.0\:\gev]_{t}$                          &$  0.00 \pm 0.00 $&$ -0.72 \pm 0.01 $\\
\noalign{\smallskip}\hline\noalign{\smallskip}
{\bf Sum}                                                 &$693.1 \pm 1.2 \pm 2.6 \pm 1.7 \pm 0.1_\psi \pm 0.7_{\rm QCD}$&$275.28 \pm 0.16 \pm 0.71 \pm 0.23 \pm 0.09_\psi  \pm 0.55_{\rm QCD}$\\
\noalign{\smallskip}\hline
\vspace{-0.2cm}
\end{tabularx}
  \caption[.]{\label{tab:results}
    Compilation of the  contributions to \amuhadLO and \dahadZ
    as obtained from HVPTools.
    Where three (or more) uncertainties are given, the first is statistical, the second
    channel-specific systematic, and the third common systematic, which 
    is correlated with at least one other channel. For the contributions 
    computed from QCD, only total uncertainties are given, which include effects
    from the $\as$ uncertainty, the truncation of the perturbative series at four loops, 
    the FOPT vs.\ CIPT ambiguity, and quark mass uncertainties. 
    Except for the latter uncertainty, all other uncertainties are taken to be 
    fully correlated among the various energy regions where QCD is used. The 
    additional uncertainty dubbed ``dual" estimates possible quark-hadron duality violating
    effects in the QCD estimate between 1.8 and 2.0$\;$GeV. The uncertainties in the Breit-Wigner 
    integrals of the narrow resonances $J/\psi$ and $\psi(2S)$ are dominated 
    by the the respective electronic width measurements~\cite{pdg10}. 
    The uncertainties in the sums (last line) are obtained by quadratically adding 
    all statistical and channel-specific systematic uncertainties, and by linearly 
    adding correlated inter-channel systematic uncertainties. 
}
\end{table*}

A compilation of the various contributions to \amuhadLO and to \dahadZ, as well as the total 
results are given in Table~\ref{tab:results}. The experimental uncertainties are separated
into statistical, channel-specific systematic, and common systematic contributions
that are correlated with at least one other channel. 

The contributions from the $J/\psi$ and $\psi(2S)$ resonances in 
Table~\ref{tab:results} are obtained by numerically integrating the corresponding 
undressed\footnote
{The undressing uses the BABAR programme {\sc Afkvac} correcting for both leptonic 
   and hadronic VP effects. The correction factors amount to 
   $(1-{\rm \Pi}(s))^2=0.956$ and $0.957$ for the $J/\psi$ and $\psi(2S)$, respectively. 
} 
Breit-Wigner lineshapes.\footnote{Using instead the narrow-width approximation, 
$\sigma_R = 12\pi^2\Gamma_{ee}^0/M_R\cdot\delta(s-M_R^2)$, gives consistent results.} 
The uncertainties in the integrals are dominated by the knowledge of the corresponding
bare electronic width $\Gamma_{R\to ee}^0$ for which we use the values $5.60 \pm 0.06\;$keV 
for $R=J/\psi$~\cite{Gamee-J/psi} and $2.35 \pm 0.05\;$keV for $R=\psi(2S)$~\cite{pdg2016}.

Sufficiently far from the quark thresholds we use four-loop~\cite{chetkuehn} 
perturbative QCD, including ${\cal O}(\as^2)$ quark mass corrections~\cite{kuhnmass}, 
to compute the inclusive hadronic cross section. Nonperturbative contributions at 
$1.8\:\gev$ were determined from data~\cite{dh98} and found to be small.
The uncertainties of the $R_{\rm QCD}$ contributions given in Table~\ref{tab:results}
are obtained from the quadratic sum of the uncertainty in $\as$  
(we use $\asZ=0.1193\pm0.0028$ from the fit
to  $Z$ precision data~\cite{gfitter}), the truncation of the perturbative 
series (we use the full four-loop contribution as systematic uncertainty), the  
difference between fixed-order perturbation theory  and, so-called, 
contour-improved perturbation theory~\cite{ledibpich}, as
well as quark mass uncertainties (we use the values and uncertainties from Ref.~\cite{pdg10}).
The former three uncertainties are taken to be fully correlated between the 
various energy regions (see Table~\ref{tab:results}), whereas the (smaller) quark-mass uncertainties are taken to be uncorrelated. 

\begin{figure}[t]
\begin{center}
\includegraphics[width=\figsize]{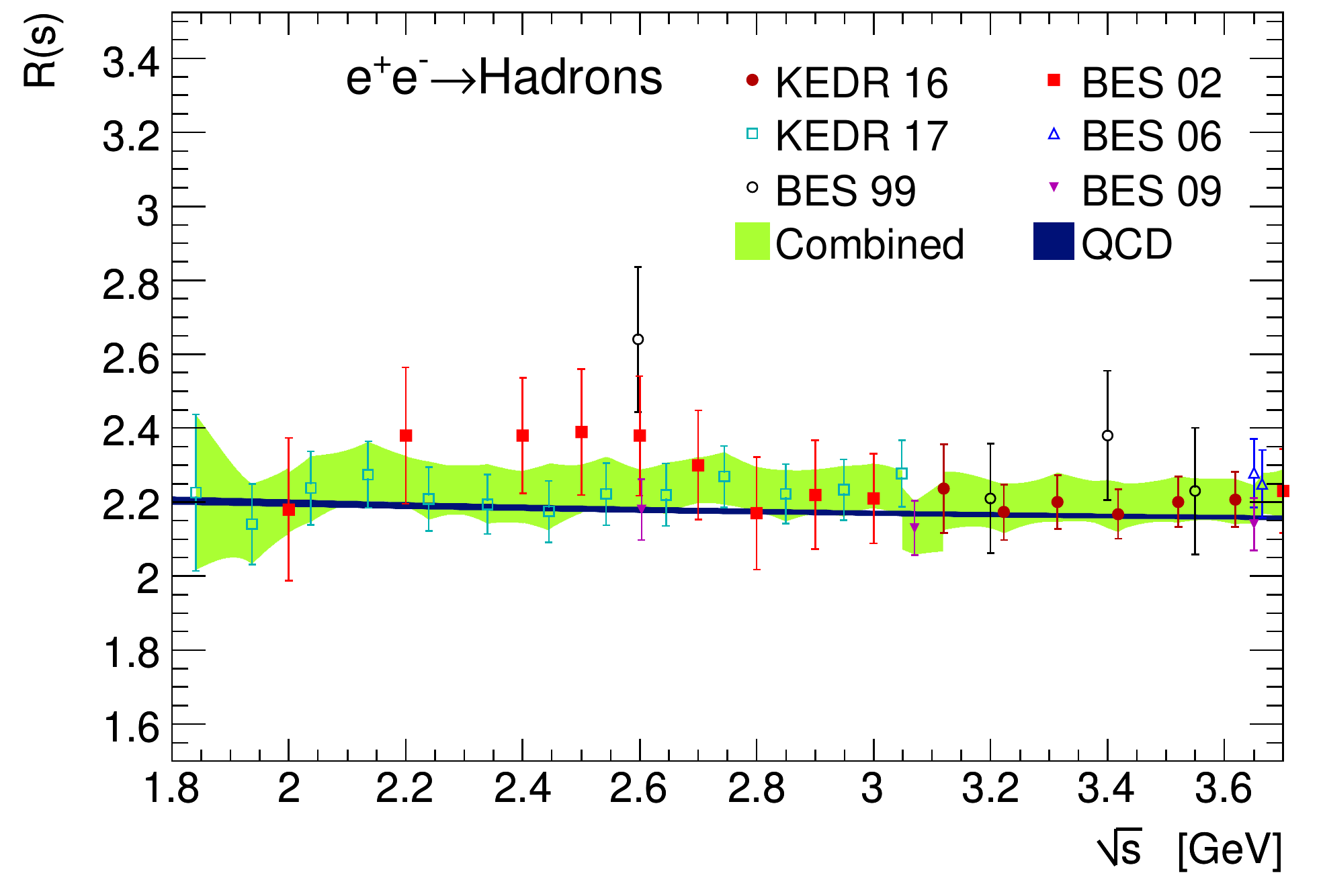}
\end{center}
\vspace{-0.2cm}
\caption[.]{ 
            Inclusive bare hadronic cross-section ratio versus centre-of-mass energy 
            in the continuum region below the $D\Dbar$ threshold. Shown are 
            BES~\cite{besR} and KEDR~\cite{kedr-r-1,kedr-r-2} 
            data points with statistical and systematic errors
            added in quadrature, the HVPTools combination (green band), and the prediction 
            from perturbative QCD (dark blue line). 
}
\label{fig:RlowE}
\end{figure}
\begin{figure*}[t]
  \centering
  \includegraphics[width=12.5cm]{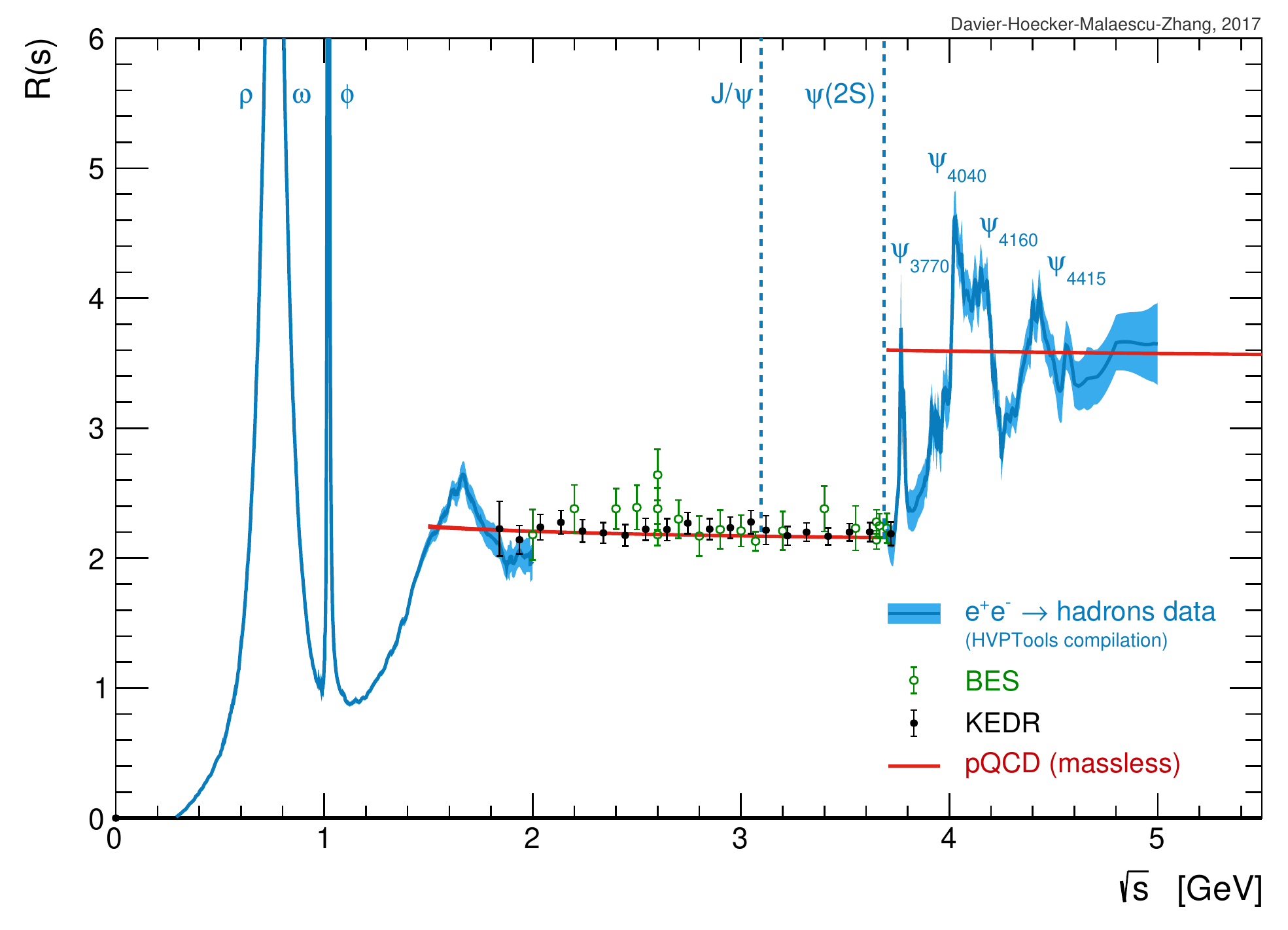}
  \vspace{-0.0cm}
  \caption{
             The total hadronic \ee annihilation rate $R$ as a function of $\sqrt{s}$. Inclusive 
             measurements from BES~\cite{besR}  
             and KEDR~\cite{kedr-r-1,kedr-r-2} are shown as data points,
             while the sum of exclusive channels from this analysis is given by the narrow 
             blue bands. Also shown is the prediction from massless perturbative QCD 
             (solid red line). 
}
  \label{fig:R}
\end{figure*}
The KEDR Collaboration has recently published results from an inclusive 
$R$ scan from $\sqrt{s}=1.84$ to 3.05$\;$GeV~\cite{kedr-r-2}, complementing their previous 
measurements obtained between 3.12 and 3.72$\;$GeV~\cite{kedr-r-1}. These data are the most 
precise and complete in this energy range with a typical systematic uncertainty
of 3\% for a total of 20 measurements. They constitute a very valuable input to 
test the validity of the perturbative QCD estimate (cf. Fig.~\ref{fig:RlowE}).
Integrating the dispersion integrals between 2.0 and 3.7$\;$GeV gives for the 
combined data $25.82 \pm 0.61$ (\amuhadLO in the usual units)
and $(21.22 \pm 0.48)\cdot10^{-4}$ (\dahadZ), compared to the QCD predictions
$25.15 \pm 0.19$ and $(20.69 \pm 0.14)\cdot10^{-4}$, respectively. 
Agreement within $1\,\sigma$ is found.

To examine the transition region between the sum of exclusive measurements and 
QCD we have computed \amuhadLO and \dahadZ in the narrow energy interval 
$1.8$--$2.0\;\gev$. For the former quantity we find $7.71  \pm 0.32$ and 
$8.30 \pm 0.09$ for data and QCD, respectively. The full difference of 
$0.59$ ($0.26\cdot10^{-4}$ in the case of \dahadZ) is assigned as additional 
systematic uncertainty, labelled by ``dual" subscripts
in Table~\ref{tab:results}. It accounts for possible low-mass quark-hadron 
duality violation affecting the perturbative QCD approximation that we 
use for this interval to avoid systematic effects due to possible unmeasured
high-multiplicity channels.

Figure~\ref{fig:R} shows the total hadronic \ee annihilation rate $R$ versus centre-of-mass 
energy as obtained from the sum of exclusive data below 2$\;$GeV and from inclusive data between 
1.8 and 5$\;$GeV.\footnote{We have verified that the integration of the finely binned $R$ 
distribution shown in Fig.~\ref{fig:R}, together with its covariance matrix, 
accurately reproduces the \amuhadLO and \dahadZ results obtained by summing the 
exclusive modes below 1.8$\;$GeV  in Table~\ref{tab:results}.} 
Also indicated are the QCD prediction above 1.5$\;$GeV and the analytical 
narrow $J/\psi$ and $\psi(2S)$ resonances.

\vspace{0.0cm}
\paragraph*{\bf\em Muon magnetic anomaly\\[0.2cm] } 

Adding all lowest-order hadronic contributions together gives
\beq
\label{eq:amuhadlo}
   \amuhadLO = 693.1 \pm 3.4\,,
\eeq
which is dominated by experimental systematic uncertainties (\cf Table~\ref{tab:results}
for a separation of the total uncertainty into its components). 
The new result is $0.4$ units 
larger than  our previous evaluation~\cite{dhmz2011} and 21\% more precise
owing to the new and improved experimental data.

Adding to~(\ref{eq:amuhadlo}) the contributions from higher order 
hadronic loops, $-9.87 \pm 0.09$ (NLO) and $1.24\pm0.01$ (NNLO)~\cite{amu-hadnlo},
hadronic light-by-light scattering, $10.5\pm 2.6$~\cite{amu-lbl}, 
as well as QED, $11\,658\,471.895 \pm 0.008$~\cite{amu-qed} (see also~\cite{pdgg-2rev} 
and references therein), and electroweak effects,
$15.36 \pm 0.10$~\cite{amu-ew}, we obtain the complete SM prediction
\beq
\label{eq:amusm}
  \amuSM = 11\,659\,182.3 \pm 3.4 \pm 2.6 \pm 0.2~(4.3_{\rm tot})\,,
\eeq
where the uncertainties account for lowest and higher order hadronic, and 
other contributions, respectively. The result~(\ref{eq:amusm}) deviates from the 
experimental value, $\amuExp=11\,659\,209.1 \pm 5.4 \pm 3.3$~\cite{bnl,pdgg-2rev}, 
by $26.8 \pm 7.6$ ($3.5\,\sigma$).

\begin{figure}[t]
\vspace{0.1cm}

\includegraphics[width=\columnwidth]{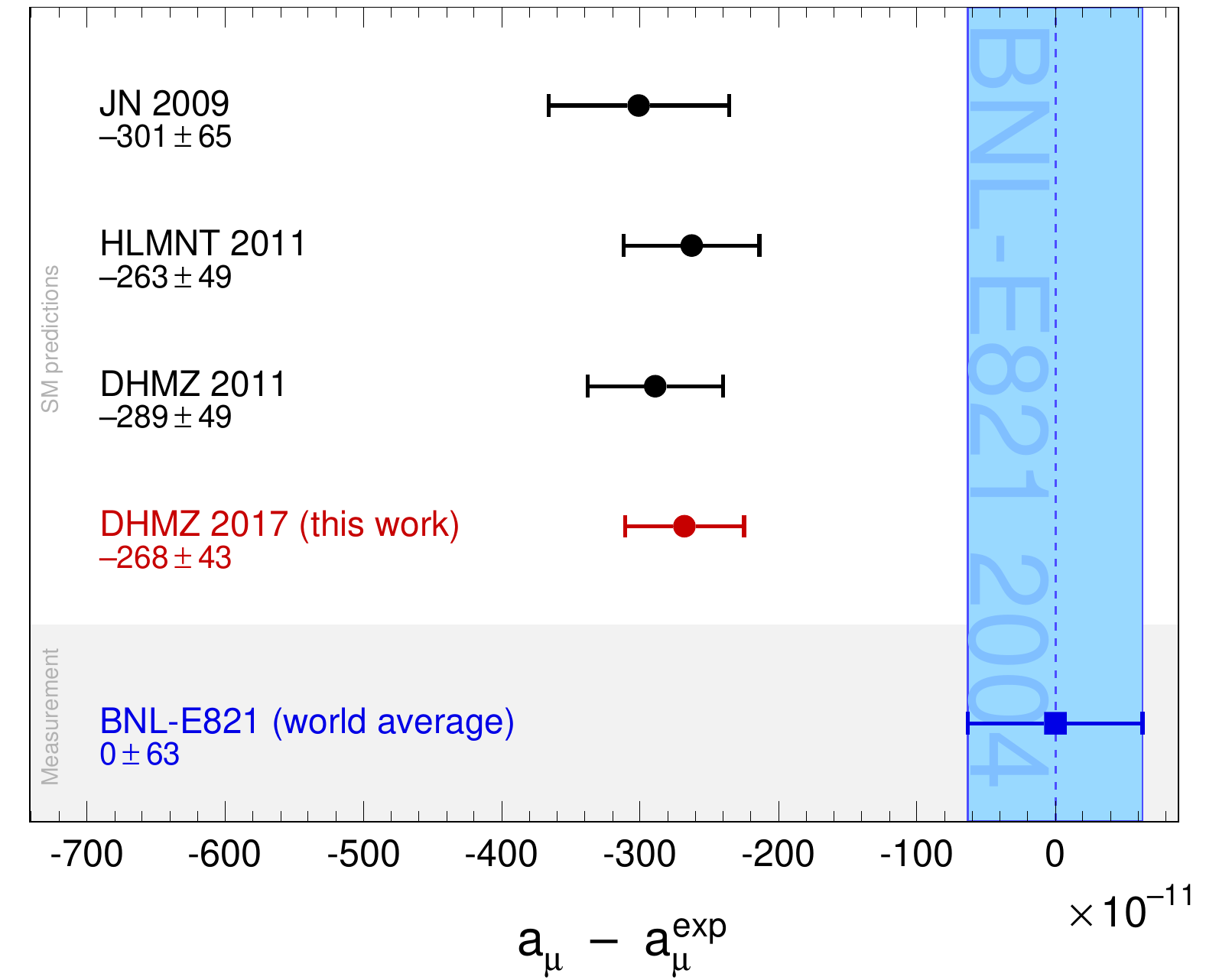}
\vspace{-0.2cm}
\caption{ 
        Compilation of recent results for $\amuSM$ (in units of $10^{-11}$),
        subtracted by the central value of the experimental average~\cite{bnl,pdgg-2rev}.
        The shaded vertical band indicates the experimental uncertainty. 
        The representative SM predictions are taken from 
        this work (DHMZ 2017), DHMZ 2011~\cite{dhmz2011}, HLMNT 2011~\cite{hlmnt}, 
        and JN 2009~\cite{jeger}. }
\label{fig:amures}
\end{figure}

A compilation of recent SM predictions for \amu compared with the experimental
result is given in Fig.~\ref{fig:amures}.

\vspace{0.0cm}
\paragraph*{\bf\em Running electromagnetic coupling at \boldmath$m_Z^2$ \\[0.2cm]}

The sum of all quark-flavour terms from Table~\ref{tab:results} gives for the 
hadronic contribution to the running of \aZ
\beq
\label{eq:dahad}
   \dahadZ   = (275.3 \pm 0.9)\cdot 10^{-4}\,,
\eeq
the uncertainty of which is dominated by data systematic effects
($0.7\cdot 10^{-4}$) and the uncertainty in the QCD prediction ($0.6\cdot 10^{-4}$).

Adding to~(\ref{eq:dahad}) the three-loop leptonic contribution,
$\Delta\alpha_{\rm lep} (m_Z^2)=314.97686\cdot 10^{-4}$~\cite{steinhauser}, 
with negligible uncertainty, one finds
\beq
   \alpha^{-1}(m_Z^2) = 128.947 \pm 0.012\,.
\eeq
The current uncertainty on $ \alpha(m_Z^2)$ is sub-dominant in the SM prediction
of the $W$-boson mass (the dominant uncertainties are due to the top mass and of 
theoretical origin), but dominates the prediction of $\sin^2\theta_{\rm eff}^\ell$, which,
however, is about twice more accurate than the combination of all 
present measurements~\cite{gfitter}.

\section{~Conclusions and perspectives}

Using newest available $e^+e^-\to {\rm hadrons}$ cross-section data
we have reevaluated the lowest-order hadronic vacuum polarisation contribution to the 
Standard Model prediction of the anomalous magnetic moment of the muon, and the hadronic contribution
to the running electromagnetic coupling strength at the $Z$-boson mass. 
For the former quantity we find $\amuhadLO = (693.1 \pm 3.4)\cdot 10^{-10}$.
The uncertainty of 0.5\% on this contribution is now reduced to about 
half the current uncertainty of the $a_\mu$ measurement, and has improved by more 
than a factor of two during the last thirteen years. The discrepancy 
between measurement and complete Standard Model prediction remains 
at a non-conclusive $3.5\,\sigma$ level. The forthcoming experiments at 
Fermilab~\cite{fnal-g-2} and JPARC~\cite{jparc-g-2},
aiming at up to four times better ultimate precision, have the potential to 
clarify the situation. 

To match the precision of these experiments  further progress is  needed to 
reduce the uncertainty on \amuhadLO from dispersion relations. 
New analyses of the dominant $\pip\pim$ channel are underway at the 
BABAR and CMD3 experiments for which a systematic uncertainty of 0.3\% 
may be reachable. In the 1--2$\;$GeV range it is important to  improve
the precision of the $\pip\pim\piz$ and $K^+ K^-$ channels.
Independently of the data-driven approach, Lattice QCD calculations 
of \amuhadLO  are  also progressing albeit not yet reaching competitive  
precision~\cite{Lattice-amu}.

The determination of \amuhadLO is closing in on the estimated  uncertainty of 
the hadronic light-by-light scattering contribution \amuhadLBL of  $2.6\cdot10^{-10}$, 
which appears irreducible at present. Here only phenomenological models have been 
used so far and Lattice QCD calculations could have a strong impact~\cite{Lattice-lbl}.

%
%


%
%

\end{document}